\documentclass[twocolumn,journal]{IEEEtran}
\usepackage[latin9]{inputenc}
\usepackage{color}
\usepackage{verbatim}
\usepackage{float}
\usepackage{amsmath}
\usepackage{amssymb}
\usepackage{graphicx}
\usepackage[numbers,numbers,sort&compress]{natbib}
\usepackage[unicode=true,
 bookmarks=true,bookmarksnumbered=true,bookmarksopen=true,bookmarksopenlevel=1,
 breaklinks=false,pdfborder={0 0 0},pdfborderstyle={},backref=false,colorlinks=true]
 {hyperref}
\hypersetup{pdftitle={Your Title},
 pdfauthor={Your Name},
 pdfpagelayout=OneColumn, pdfnewwindow=true, pdfstartview=XYZ, plainpages=false}

\makeatletter
\let\oldforeign@language\foreign@language
\DeclareRobustCommand{\foreign@language}[1]{%
  \lowercase{\oldforeign@language{#1}}}

\usepackage[ruled,vlined,linesnumbered]{algorithm2e}

\usepackage{graphicx}

\SetKwInput{KwInput}{Input}
\SetKwInput{KwOutput}{Output}

\usepackage{algorithmicx}
\usepackage{algpseudocode}
\usepackage{amsmath,amsfonts,amssymb}

\usepackage{array,booktabs,arydshln}
\usepackage{xcolor}
\usepackage{arydshln}

\usepackage{float} 

\usepackage[caption=false,font=footnotesize]{subfig}

 \hypersetup{
     bookmarks=true,         
     unicode=false,          
     pdftoolbar=true,        
     pdfmenubar=true,        
     pdffitwindow=false,     
     pdfstartview={FitH},    
     pdftitle={My title},    
     pdfauthor={Author},     
     pdfsubject={Subject},   
     pdfcreator={Creator},   
     pdfproducer={Producer}, 
     pdfkeywords={keyword1, key2, key3}, 
     pdfnewwindow=true,      
     colorlinks=true,       
     linkcolor=magenta,          
     citecolor=magenta,        
     filecolor=magenta,      
     urlcolor=magenta           
 }

\usepackage{tikz}

\usepackage{tcolorbox}

\@ifundefined{showcaptionsetup}{}{%
 \PassOptionsToPackage{caption=false}{subfig}}
\usepackage{subfig}
\makeatother

\begin{document}
\title{Uncertainty-Aware Resource Provisioning \\for Network Slicing}
\author{Quang-Trung~Luu, Sylvaine~Kerboeuf, and~Michel~Kieffer \thanks{Q.-T.~Luu is with Nokia Bell Labs and the L2S, CNRS-CentraleSup\'elec-Univ
Paris-Sud-Univ Paris-Saclay, France, e-mail: quang\_ trung.luu@nokia.com.}\thanks{S.~Kerboeuf is with Nokia Bell Labs, France, e-mail: sylvaine.kerboeuf@
nokia-bell-labs.com.}\thanks{M.~Kieffer is with the L2S, CNRS-CentraleSup\'elec-Univ Paris-Sud-Univ
Paris-Saclay, France, e-mail: michel.kieffer@l2s.centralesupelec.fr.}}
\markboth{Preprint}{Q.-T.~Luu \MakeLowercase{\emph{et al.}}: Your Title}
\maketitle
\begin{abstract}
Network slicing allows Mobile Network Operators to split the physical
infrastructure into isolated virtual networks (slices), managed by
Service Providers to accommodate customized services. The Service
Function Chains (SFCs) belonging to a slice are usually deployed on
a best-effort premise: nothing guarantees that network infrastructure
resources will be sufficient to support a varying number of users,
each with uncertain requirements.

Taking the perspective of a network Infrastructure Provider (InP),
this paper proposes a resource provisioning approach for slices, robust
to a partly unknown number of users with random usage of the slice
resources. The provisioning scheme aims to maximize the total earnings
of the InP, while providing a probabilistic guarantee that the amount
of provisioned network resources will meet the slice requirements.
Moreover, the proposed provisioning approach is performed so as to
limit its impact on low-priority background services, which may co-exist
with slices in the infrastructure network. 

A Mixed Integer Linear Programming formulation of the slice resource
provisioning problem is proposed. Optimal joint and suboptimal sequential
solutions are proposed. These solutions are compared to a provisioning
scheme that does not account for best-effort services sharing the
common infrastructure network.

\end{abstract}


\IEEEpeerreviewmaketitle{}

\section{Introduction}

Network slicing will play an essential role in 5G communication systems
\cite{5GAmericas2016,IETF2017,Barakabitze2020}. Leveraging Network
Function Virtualization (NFV), network slicing reduces overall equipment
and management costs \cite{Liang2014} by increasing flexibility in
the way the network is operated \cite{Rost2017}. Multiple dedicated
end-to-end virtual networks or \emph{slices} can be managed in parallel
over a given infrastructure network. With network slicing, vertical
markets can be addressed: Customers can manage their own applications
by exploiting built-in network slices tailored to their needs \cite{GSMA2017}.

In the extended survey \cite{Barakabitze2020} of the so far research
efforts in 5G network slicing, the authors provide a taxonomy of network
slicing, architectures and future challenges. One of the significant
questions is how to meet the slice requirements of different verticals,
where multiple network segments including the radio access, transport,
and core networks, have to be considered. Infrastructure networks
on which slices are operated must support high-quality services with
increasing resource consumption (video streaming, telepresence, augmented
reality, remote vehicle operation, gaming, \textit{etc}.). Moreover,
the number of users of each slice, their location (usually difficult
to predict \cite{Richart2016}), and resource demands may fluctuate
with time. These uncertainties may impact significantly the resources
consumed by each slice and raise the challenging problem of \textit{slice
resource} \emph{provisioning}. Enough infrastructure resources should
be dedicated to a given slice to ensure an appropriate Quality of
Service (QoS) despite the uncertainties in the number of slice users
and their demands. Over-provisioning should also be avoided, to limit
the infrastructure leasing costs and leave resources to concurrent
slices.

Existing work on network slicing, see, \textit{e.g}., \cite{Huin2017,Wang2017,Su2019,Barakabitze2020},
is mainly focused on the resource allocation aspect, \textit{i.e}.,
assigning infrastructure network resources to virtual network components,
with the aim to maximize resource utilization and minimize operation
costs. The traffic dynamics in individual slices, such as flow arrival/departure,
as well as the dynamics of resource availability on the network infrastructure,
may lead to slice QoS below the level expected by the Service Provider
(SP) managing the slice. Consequently, to fully unleash the power
of network slicing in dynamic environments, uncertainties related
to the resource demands need to be carefully addressed.

This paper investigates a method to provision infrastructure resources
for network slices, while being robust to a partly unknown number
of users with a random usage of the slice resources. Moreover, since
some parts of the infrastructure network on which slices should be
deployed are often already employed by low-priority background services,
the provisioning approach will be performed so as to limit its impact
on these services.

The rest of the paper is structured as follows. Section~\ref{sec:Related-work}
analyzes related work, and highlights our main contributions. The
model of the infrastructure network and of the slice resource demands
are presented in Section~\ref{sec:Problem-Description}. The robust
slice resource provisioning problem with uncertainties in the number
of users as well as in the resource demands and accounting for the
best-effort background services is then formulated in Section~\ref{sec:Optimal-Solution}.
The robust slice provisioning problem is transformed into a mixed
integer linear programming (MILP) problem in Section~\ref{sec:Suboptimal-Solution}.
Numerical results are presented in Section~\ref{sec:Evaluation}.
Finally, Section~\ref{sec:Conclusions} draws some conclusions and
perspectives.

\section{Related work \label{sec:Related-work}}

Several works on uncertainty-aware resource allocation for virtual
networks can be found in the literature. 

In many conventional approaches enough network resources are allocated
to make a service available to all users, all the time. In \cite{Trinh2011},
flexible service availability levels are defined, leading to cost
savings for the infrastructure provider that can offer overbooked
resources for users accepting a service with possibly degraded availability.
In the context of network slicing, SPs can benefit from such an approach
by providing services with reduced availability or degraded quality
to some users ready to accept these conditions. Nevertheless, to evaluate
the incidence on the QoS of such under-provisioning mechanism, it
is necessary to introduce models of the number of users of a service
and of the resource consumption, not considered in \cite{Trinh2011}.

A worst-case allocation at peak traffic is considered in \cite{Huin2017,Wang2017}.
Nevertheless, this infrastructure resource overbooking is costly and
most of the time unnecessary, as all individual slice resource demands
are very unlikely peaking simultaneously. In \cite{Coniglio2015},
the virtual network embedding problem is solved considering uncertain
traffic demands. An MILP formulation is considered, where some of
the constraints are required to be satisfied with high probability.
In \cite{Mireslami2019}, the total deployment costs for cloud computing
applications are minimized, while satisfying some QoS constraints.
To cope with the uncertain nature of the demands, a stochastic optimization
approach is adopted by modeling user demands as random variables obeying
normal distributions. Deployment is performed based on the mean demands
increased by an integer amount of their standard deviations. This
might lead to a conservative solution, requiring more allocated resources
than needed. This also reduces somehow the possibility of having service-dependent
required confidence levels.

A network slice embedding problem is considered in \cite{Fendt2019},
where available resources and resource demands are assumed to be partly
uncertain. They are described by normal distributions built upon the
data history on mobile network resource availability as well as slice
resource utilization. To control the probability that a slice embedding
solution will benefit from enough infrastructure resource, despite
the uncertainties, some adjustable safety factor $\gamma$ is introduced.
As in \cite{Mireslami2019}, enough resources are dedicated to a service
so as to satisfy the mean plus $\gamma$ times the standard deviation
of the demands. In \cite{Fendt2019}, additionally, a similar approach
is considered to account for the uncertainty in the available resources.
A \emph{probability of feasibility}, depending on $\gamma$, is then
evaluated for the slice embedding to measure the risk of having a
degraded service for some users. The proposed solution leads to a
slice resource allocation solution robust to uncertainties. Nevertheless,
the resource demands of the different components of the slice have
been considered as independent. Moreover, the safety factor $\gamma$
is chosen identical for resource demands and available resources.
This again may lead to allocating more resources than strictly necessary,
and increases the operation cost. 

The network slice embedding problem with demand uncertainties is also
addressed in \cite{Baumgartner2018}. The minimization of deployment
costs considering first static resource demands is formulated as an
MILP. Two robust network slice design formulations are then proposed,
in uncorrelated and correlated demand uncertainties are considered.
In both cases, the objective function is unchanged but some constraints
become nonlinear due to the addition of inner maximization problems.
These problems account for the upper bound of the resource demands,
thus making the network slice embedding problem more complex. A linearization
technique inspired by \cite{Bertsimas2003} is proposed to relax these
inner problems. A tuning parameter $\Gamma$ is introduced to control
the trade-off between robustness to the demand uncertainties and the
deployment costs. Uncertainties related to the background traffics
on the infrastructure, which clearly affect the residual infrastructure
resources, are not considered.

To reduce the computation effort required to solve the robust network
slice embedding problem, \cite{Bauschert2019} proposes to use a genetic
algorithm, shown to surpass the performance of state-of-the-art robust
MILP solvers used, \emph{e.g.}, in \cite{Baumgartner2018}. Uncertainties
in infrastructure link bandwidth are also considered in \cite{Wen2019},
where possible failures of infrastructure nodes or links are taken
into account to propose a robust algorithm that minimizes the network
resource consumption under uncertain demands, while remapping the
network slice in case of infrastructure failures. Since \cite{Baumgartner2018},
\cite{Bauschert2019}, and \cite{Wen2019} assume that the distribution
of the variable demands and available infrastructure resource are
unknown, their optimization are relatively conservative. Furthermore,
uncertainties in various types of resources such as computing, memory,
or wireless are not addressed.

In all the above works, the effect of the best effort background services
combined with a approach robust to uncertainties in the demands and
in the infrastructure resources has not yet been considered for the
slice provisioning problem. As shown in the sequel these are two important
aspects that need to be taken into account for efficiently providing
slices with guaranteed Service Level Agreement (SLA). Finally, we
emphasize that these approaches are solving the problem of resource
allocation rather than provisioning, \emph{i.e.}, reserving infrastructure
resource for a further allocation.

In this paper, we adopt the point of view of the Infrastructure network
Provider (InP). We propose a provisioning scheme which aims at maximizing
the total earnings of the InP, while providing a probabilistic guarantee
that the amount of provisioned network resources will meet the slice
requirements. In the provisioning approach, various infrastructure
network resources are booked for a slice to satisfy its requirements.
Slice resource demands are aggregated. Consequently, resources of
several infrastructure nodes may have to be gathered and parallel
physical links have to be considered to satisfy these aggregated demands.
The provisioning approach may be performed prior to the resource allocation
at the time of deployment described, \emph{e.g.}, in \cite{Riggio2016,Vizarreta2017},
where virtual nodes and links are mapped on the infrastructure network.
Moreover, instead of considering uncertainties in the available network
resource, as in \cite{Fendt2019}, here, we consider best-effort background
services running in parallel with the network slices on the infrastructure
network. The proposed scheme is able to maintain the impact of resource
provisioning on those background services at a prescribed level. Previous
results on slice resource provisioning have been presented in \cite{Luu2020b}.
Nevertheless, uncertainties in the number of users of a slice and
in the way they consume resources, as well as concurrent best-effort
services sharing the infrastructure network have not been taken into
account.

\section{Notations and Hypotheses \label{sec:Problem-Description}}

\begin{figure}[tbh]
\begin{centering}
\includegraphics[width=0.7\columnwidth]{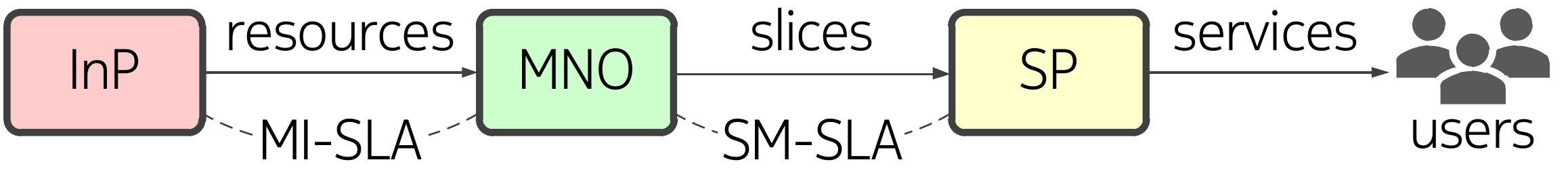}
\par\end{centering}
\caption{Network slicing entities and their SLA-based relationships\label{fig:System-architecture}}
\end{figure}

A typical network slicing system involves several entities. This may
include one or many InPs, Mobile Network Operators (MNOs), and SPs,
as depicted in Figure~\ref{fig:System-architecture} \cite{Liang2014}.
An InP owns and manages the wireless and wired infrastructure such
as the cell sites, the fronthaul and backhaul networks, and cloud
data centers. An MNO leases resources from InPs to setup and manage
the slices. An SP then exploits the slices supplied by an MNO, and
provides to its customers the required services that are running within
the slices. Service needs are forwarded by an SP to an MNO within
an SLA denoted SM-SLA in what follows.

The SM-SLA describes, at a high level of abstraction, characteristics
of the service with the desired QoS. These characteristics may be
time-varying due, \emph{e.g.}, to user mobility. In this paper, one
considers SM-SLAs composed of: \textit{i}) a probability mass function
(pmf) describing the target number of users/devices to be supported
by the slice, \textit{ii}) a description of the characteristics of
the service and of the way it is employed by a typical user/device,
and \textit{iii}) a target probability of service satisfaction. In
addition, several time intervals may be considered in the SM-SLA,
intervals over each of which the service characteristics and constraints
are assumed constant, but may vary from one interval to the next one.
These time intervals translate,\textit{ e.g}., day and night variations
of user demands. They last between tens of minutes to hours. It is
of the responsibility of the SP and MNO to properly scale the requirements
expressed in the SM-SLA, by considering, for example, similar services
deployed in the past.

Taking the InP perspective, our aim, with resource provisioning is
to reserve, somewhat in advance, enough infrastructure resources to
ensure that the MNO will be able to provide a slice with characteristics
as stated in the SM-SLA it has with the SP. The time scale at which
provisioning is performed is much larger than that at which slices
are deployed and adapted to meet actual time-varying user demands.
In what follows, one focuses on a given time interval over which resources
will be provisioned so as to be compliant with the variations of user
demands within a slice. The duration of this time interval results
from a compromise between the need to update the provisioning and
the level of conservatism in the amount of provisioned resources required
to satisfy fast fluctuating user demands.

Each slice consists of one or multiple Service Function Chains (SFCs)
of different types. An SFC consists of an ordered set of interconnected
Virtual Network Functions\textit{\emph{ (}}VNFs) describing the processing
applied to data flows related to a given service. The MNO translates
the SP high-level demands into SFCs able to fulfill the service requirements.
Based on the characteristics of the service and of its usage, the
MNO describes the way the slice (SFCs) resources are consumed by a
given user/device. To characterize the variability over time and among
users of these demands, we assume that the MNO considers a probabilistic
description of the consumption of slice resources by a typical user.
The MNO then forwards to the InP these characteristics as part of
an SLA between them (MI-SLA). Each InP then provisions infrastructure
resources needed for the SFCs. Under the MI-SLA, this provisioning
has to meet the target probability of service satisfaction. This translates
the fact that enough resources of various types have been provisioned
to satisfy the resource demands of the users of the service. This
probability is evaluated considering the pmf describing the number
of users of the service and the probabilistic description of the slice
resource consumption by a typical user. When performing the provisioning,
each InP has to limit the impact on other best-effort service running
on its infrastructure network.

In this paper, one considers an infrastructure owned by a single InP.
To perform the provisioning, the InP has to identify the infrastructure
nodes which will provide resources for future deployment of VNFs and
the links able to transmit data between these nodes, while respecting
the structure of SFCs and optimizing a given objective (\textit{e.g}.,
minimizing the infrastructure and software fee costs).

Table~\ref{tab:notations} summarizes all parameters involved in
the description of the infrastructure network and the graph of SFCs
for a slice.

\begin{table}[tbh]
\caption{Table of Notations\label{tab:notations}}

\centering
\begin{tabular}{
p{0.05\columnwidth}
p{0.3\columnwidth} }
\toprule

\multicolumn{1}{c}{ \textit{Symbol} }

& \multicolumn{1}{l}{ \textit{Description}}\\

\cmidrule[0.4pt](lr{0.12em}){1-1}%
\cmidrule[0.4pt](lr{0.12em}){2-2}%

\multicolumn{1}{c}{ $\mathcal{G}$ } & \multicolumn{1}{l}{ Infrastructure
network graph, $\mathcal{G}=\left(\mathcal{N},\mathcal{E}\right)$
}\\

\multicolumn{1}{c}{ $\mathcal{N}$ } & \multicolumn{1}{l}{ Set of
infrastructure nodes }\\

\multicolumn{1}{c}{ $\mathcal{E}$ } & \multicolumn{1}{l}{ Set of
infrastructure links }\\

\multicolumn{1}{c}{ $a_{n}\left(i\right)$ } & \multicolumn{1}{l}{
Available resource of type $n$ at node~$i$}\\

\multicolumn{1}{c}{ $a_{\text{b}}\left(ij\right)$ } & \multicolumn{1}{l}{
Available bandwidth of link~$ij$ }\\

\multicolumn{1}{c}{ $c_{n}\left(i\right)$ } & \multicolumn{1}{l}{
Per-unit cost of resource of type $n$ for node~$i$ }\\

\multicolumn{1}{c}{ $c_{\text{b}}\left(ij\right)$ } & \multicolumn{1}{l}{
Per-unit cost for link~$ij$ }\\

\multicolumn{1}{c}{ $c_{\text{f}}\left(i\right)$ } & \multicolumn{1}{l}{
Fixed cost for using node~$i$}\\

\multicolumn{1}{c}{ $\mathcal{S}$ } & \multicolumn{1}{l}{ Set of
slices to be deployed }\\

\multicolumn{1}{c}{ $\mathcal{G}_{s}$ } & \multicolumn{1}{l}{ SFC
graph, $\mathcal{G}_{s}=\left(\mathcal{N}_{s},\mathcal{E}_{s}\right)$
}\\

\multicolumn{1}{c}{ $\mathcal{N}_{s}$ } & \multicolumn{1}{l}{ Set
of VNFs~$v$ }\\

\multicolumn{1}{c}{ $\mathcal{E}_{s}$ } & \multicolumn{1}{l}{ Set
of interconnections~$vw$ between VNF~$v$ and~$w$}\\

\multicolumn{1}{c}{ $r_{s,n}\left(v\right)$ } & \multicolumn{1}{l}{
Fixed amount of resources of type $n$ required }\\

\multicolumn{1}{c}{ } & \multicolumn{1}{l}{ by an instance of VNF~$v$
to operate properly}\\

\multicolumn{1}{c}{ $r_{s,\text{b}}\left(vw\right)$ } & \multicolumn{1}{l}{
Fixed amount of bandwidth to sustain traffic }\\

\multicolumn{1}{c}{ } & \multicolumn{1}{l}{ demand between VNF instances~$v$
and~$w$}\\

\multicolumn{1}{c}{ $U_{s,n}\left(v\right)$ } & \multicolumn{1}{l}{
Random amount of resources of type $n$ }\\

\multicolumn{1}{c}{ } & \multicolumn{1}{l}{ of virtual node~$v$
employed by a user }\\

\multicolumn{1}{c}{ $U_{s,\text{b}}\left(vw\right)$ } & \multicolumn{1}{l}{
Random amount of bandwidth of virtual link~$vw$ }\\

\multicolumn{1}{c}{ } & \multicolumn{1}{l}{ employed by a user }\\

\multicolumn{1}{c}{ $R_{s,n}\left(v\right)$ } & \multicolumn{1}{l}{
Random amount of resources of type $n$ }\\

\multicolumn{1}{c}{ } & \multicolumn{1}{l}{ of virtual node~$v$
employed by $N_{s}$ users }\\

\multicolumn{1}{c}{ $R_{s,\text{b}}\left(vw\right)$ } & \multicolumn{1}{l}{
Random amount of bandwidth of virtual link~$vw$ }\\

\multicolumn{1}{c}{ } & \multicolumn{1}{l}{ employed by $N_{s}$
users }\\

\multicolumn{1}{c}{ $B_{s,n}\left(i\right)$ } & \multicolumn{1}{l}{
Amount of resources of type $n$ on infrastructure }\\

\multicolumn{1}{c}{ } & \multicolumn{1}{l}{ node~$i$ consumed
by background services }\\

\multicolumn{1}{c}{ $B_{s,\text{b}}\left(ij\right)$ } & \multicolumn{1}{l}{
Amount of bandwidth on infrastructure }\\

\multicolumn{1}{c}{ } & \multicolumn{1}{l}{ link~$ij$ consumed
by background services }\\

\bottomrule
\end{tabular}
\end{table}

\subsection{Infrastructure Network\label{subsec:Infrastructure-Network}}

Consider an infrastructure network managed by a given InP. This network
is represented by a directed graph $\mathcal{G}=\left(\mathcal{N},\mathcal{E}\right)$,
where $\mathcal{N}$ is the set of infrastructure nodes and $\mathcal{E}$
is the set of infrastructure links, which correspond to the wired
connections between and within nodes (loopback links) of the infrastructure
network.

Each infrastructure node $i\in\mathcal{N}$ is characterized by a
given amount of available computing, memory, and wireless resources,
denoted as $a_{\text{c}}(i)$, $a_{\text{m}}(i)$, and $a_{\text{w}}\left(i\right)$,
which may be \textit{allocated} to new network slices. These amounts
correspond to the total available resources reduced by the amount
of resources previously provisioned to concurrent slices. An operation
cost paid by the InP is attributed to each unit of node resource.
The per-unit node resource cost associated to a given node~$i$ consists
of a fixed part $c_{\text{f}}\left(i\right)$ for node disposal (paid
for each slice using node~$i$), and variable parts $c_{\text{c}}(i)$,
$c_{\text{m}}(i)$, and $c_{\text{w}}(i)$, which depend linearly
on the amount of resources provided by that node.

Similarly, each infrastructure link $ij\in\mathcal{E}$ connecting
node $i$ to $j$ has an available bandwidth $a_{\text{b}}\left(ij\right)$,
and an associated per-unit bandwidth cost $c_{\text{b}}(ij)$. Several
distinct VNFs of the same slice may be deployed on a given infrastructure
node. When communication between these VNFs is required, an internal
(loopback) infrastructure link $ii\in\mathcal{E}$ can be used at
each node $i\in\mathcal{N}$, as in \cite{Wang2009}, in the case
of interconnected virtual machines (VMs) deployed on the same host.
The associated per-unit bandwidth cost, in that case, is $c_{\textrm{\text{b}}}\left(ii\right)$.

\subsection{Graphs of Resource Demands\label{subsec:Graphs-of-Resource-Demands}}

A demand of resources is defined on the basis of an SLA between an
SP and the MNO. As in \cite{Luu2020b}, we consider that a slice is
devoted to a single type of service supplied by a given type of SFC.
Several instances of that SFC may have to be deployed so as to satisfy
the user demand. The topology of each SFC of slice~$s$ is represented
by a graph $\mathcal{G}_{s}=\left(\mathcal{N}_{s},\mathcal{E}_{s}\right)$
representing the VNFs and their interconnections. Each virtual node
$v\in\mathcal{N}_{s}$ represents an instance of a VNF, and each virtual
link $vw\in\mathcal{E}_{s}$ represents the connection between virtual
nodes~$v$ and $w$.

The following \textit{weighted }graphs are build upon $\mathcal{G}_{s}$.
\begin{itemize}
\item $\mathcal{G}_{s}^{\text{r}}=\left(\mathcal{N}_{s}^{\text{r}},\mathcal{E}_{s}^{\text{r}}\right)$
is the graph of Resource Demands of an SFC (SFC-RD) of slice~$s$.
Each node $v\in\mathcal{N}_{s}^{\text{r}}$ is characterized by a
fixed amount of computing $r_{s,\text{c}}(v)$ and memory $r_{s,\text{m}}(v)$
resources allocated by the infrastructure node on which the VNF instance~$v$
is deployed to operate properly. Each link~$vw\in\mathcal{E}_{s}^{\text{r}}$
is characterized by a given amount of bandwidth $r_{s,\text{b}}(vw)$
that has to be allocated by the infrastructure network to sustain
the traffic demand between VNF instances~$v$ and~$w$.
\item $\mathcal{G}_{s}^{\text{U}}=\left(\mathcal{N}_{s}^{\text{U}},\mathcal{E}_{s}^{\text{U}}\right)$
is the graph of Resource Demands a typical User (U-RD) of slice~$s$.
Each user of slice~$s$ is assumed to consume a random proportion
of the resources of an SFC of that slice. In addition, the consumed
resources by various users are represented by independently and identically
distributed random vectors. For a typical user, let $U_{s,\text{c}}\left(v\right)$,
$U_{s,\text{m}}\left(v\right)$, $U_{s.\text{w}}\left(v\right)$,
and $U_{s,\text{b}}\left(vw\right)$ be the random amount of employed
resources of VNF instance~$v\in\mathcal{N}_{s}^{\text{r}}$ and of
virtual link $vw\in\mathcal{E}_{s}^{\text{r}}$ of the SFC-RD $\mathcal{G}_{s}^{\text{r}}$.
\item $\mathcal{G}_{s}^{\text{R}}=\left(\mathcal{N}_{s}^{\text{R}},\mathcal{E}_{s}^{\text{R}}\right)$
is the graph of Resource Demands of Slice~$s$ (S-RD). The weight
of each node~$v\in\mathcal{N}_{s}^{\text{R}}$ and of each link~$vw\in\mathcal{E}_{s}^{\text{R}}$
represents the aggregate amount of resources employed by a random
number $N^{s}$ of independent users of slice~$s$. These amounts
are described by random variables denoted as $R_{s,\text{c}}\left(v\right)$,
$R_{s,\text{m}}\left(v\right)$, $R_{s,\text{w}}\left(v\right)$,
and $R_{s,\text{b}}\left(vw\right)$, for computing, memory, wireless,
and bandwidth demand, respectively.
\end{itemize}
Considering the analysis of co-allocated online services of large
scale data centers reported in \cite{Jiang2019}, the utilization
of CPU and memory of virtual machines (VMs) have a positive correlation
in the majority of cases. Moreover, this correlation is particularly
strong at the VMs that execute the same jobs, showing correlation
coefficients larger than $0.85$. Based on this observation, for a
typical user, the resource demands of different types for a given
node $v\in\mathcal{N}_{s}^{\text{U}}$ are considered to be correlated.
The demands for resources of the same type among virtual nodes are
also correlated. Finally, the resulting traffic demands between nodes
is usually also correlated with the resource demands for a given virtual
node. To represent this correlation, consider the vector of joint
resource demands for a typical user of an SFC of slice~$s$
\[
\mathbf{U}_{s}=\left(U_{s,\text{c}}\left(v\right),U_{s,\text{m}}\left(v\right),U_{s,\text{w}}\left(v\right),U_{s,\text{b}}\left(vw\right)\right)_{\left(v,vw\right)\in\mathcal{G}_{s}^{\text{U}}}^{\top}.
\]
Assuming that $U_{s,\text{c}}\left(v\right)$, $U_{s,\text{m}}\left(v\right)$,
$U_{s.\text{w}}\left(v\right)$, and $U_{s,\text{b}}\left(vw\right)$
are normally distributed, $\mathbf{U}_{s}$ follows a multivariate
normal distribution with probability density
\begin{equation}
f\left(\mathbf{x};\boldsymbol{\mu}_{s},\boldsymbol{\Gamma}_{s}\right)=\frac{1}{\sqrt{\left(2\pi\right)^{\textrm{card}\left(\mathbf{U}_{s}\right)}\left|\boldsymbol{\Gamma}_{s}\right|}}e^{-\frac{1}{2}\left(\mathbf{x}-\boldsymbol{\mu}_{s}\right)^{\top}\left(\boldsymbol{\Gamma}_{s}\right)^{-1}\left(\mathbf{x}-\boldsymbol{\mu}_{s}\right)},\label{eq:JointResDist}
\end{equation}
with mean
\[
\boldsymbol{\mu}_{s}=\left(\mu_{s,\text{c}}\left(v\right),\mu_{s,\text{m}}\left(v\right),\mu_{s,\text{w}}\left(v\right),\mu_{s,\text{b}}\left(vw\right)\right)_{\left(v,vw\right)\in\mathcal{G}_{s}^{\text{U}}}^{\top},
\]
and covariance matrix $\boldsymbol{\Gamma}_{s}$ such that
\[
\text{diag}\left(\boldsymbol{\Gamma}_{s}\right)=\left(\sigma_{s,\text{c}}^{2}\left(v\right),\sigma_{s,\text{m}}^{2}\left(v\right),\sigma_{s,\text{w}}^{2}\left(v\right),\sigma_{s,\text{b}}^{2}\left(vw\right)\right)_{\left(v,vw\right)\in\mathcal{G}_{s}^{\text{U}}}^{\top},
\]
the off-diagonal elements of $\boldsymbol{\Gamma}_{s}$ representing
the correlation between different types of resource demands. In \eqref{eq:JointResDist},
$\textrm{card}\left(\mathbf{U}_{s}\right)$ is the number of elements
of $\mathbf{U}_{s}$. One has thus $U_{s,n}(v)\sim\mathcal{N}\left(\mu_{s,n}\left(v\right),\sigma_{s,n}^{2}\left(v\right)\right)$,
with $n\in\left\{ \text{c},\text{m},\text{w}\right\} $ and $U_{s,\text{b}}\left(vw\right)\sim\mathcal{N}\left(\mu_{s,\text{b}}\left(vw\right),\sigma_{s,\text{b}}^{2}\left(vw\right)\right)$.

Assume that the number of users $N_{s}$ to be supported by slice~$s$
is described by the pmf 
\begin{equation}
p_{k}=\Pr\left(N_{s}=k\right).\label{eq:UserDist}
\end{equation}
Since the amount of resources of VNF~$v$ and of virtual link~$vw$
consumed by different users is represented by independently and identically
distributed copies of $\mathbf{U}_{s}$, the joint distribution of
the aggregate amount $\mathbf{U}_{s,k}$ of resources consumed by
$k$ independent users is $f\left(\mathbf{x},k\boldsymbol{\mu}_{s},k^{2}\boldsymbol{\Gamma}_{s}\right)$.
The total amount of resources employed by a random number $N_{s}$
of independent users, $\mathbf{R}_{s}=\mathbf{U}_{s,N_{s}}=\left(R_{s,\text{c}}\left(v\right),R_{s,\text{m}}\left(v\right),R_{s,\text{w}}\left(v\right),R_{s,\text{b}}\left(vw\right)\right)_{\left(v,vw\right)\in\mathcal{G}_{s}^{\text{R}}}^{\top}$
, is distributed according to
\begin{equation}
g\left(\mathbf{x},\boldsymbol{\mu}_{s},\boldsymbol{\Gamma}_{s}\right)=\sum_{k=0}^{\infty}p_{k}f\left(\mathbf{x},k\boldsymbol{\mu}_{s},k^{2}\boldsymbol{\Gamma}_{s}\right).\label{eq:PDF_Rs}
\end{equation}

\begin{figure}[tbh]
\begin{centering}
\subfloat[Uncorrelated demands\label{fig:Prob_Ex_Uncorr}]{\begin{centering}
\includegraphics[width=0.48\columnwidth]{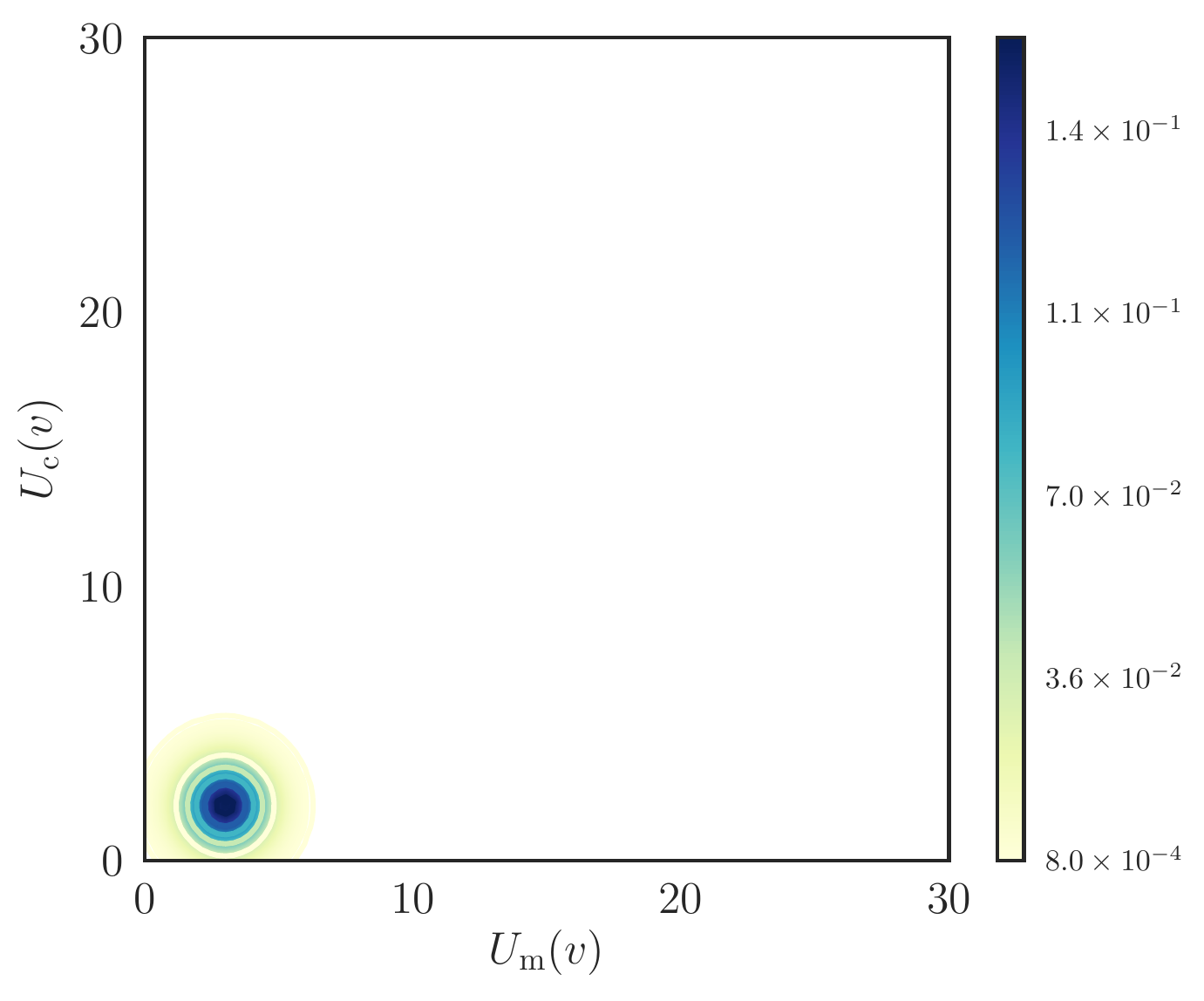}\includegraphics[width=0.48\columnwidth]{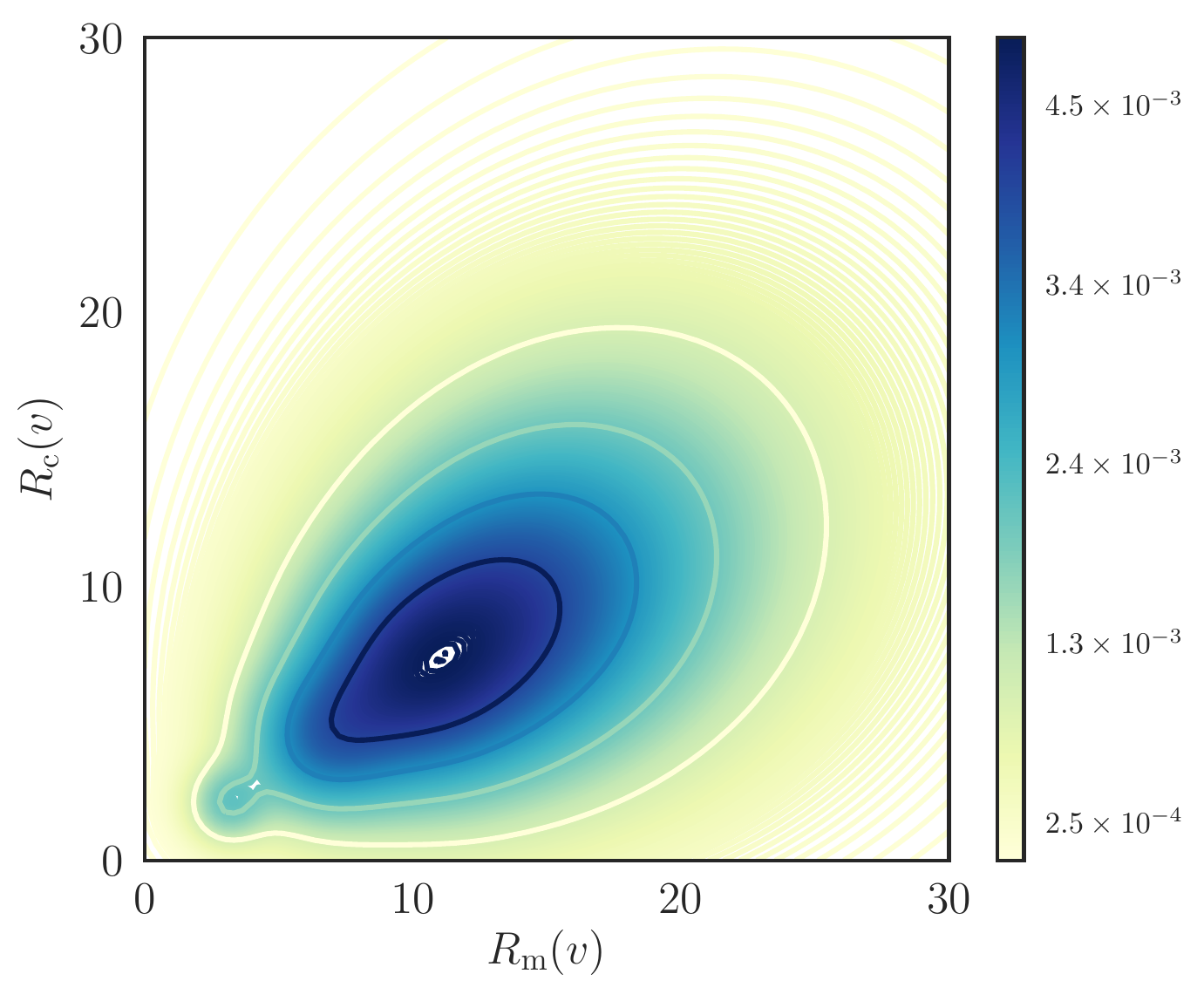}
\par\end{centering}
}
\par\end{centering}
\begin{centering}
\subfloat[Correlated demands\label{fig:Prob_Ex_Un_Corr}]{\begin{centering}
\includegraphics[width=0.48\columnwidth]{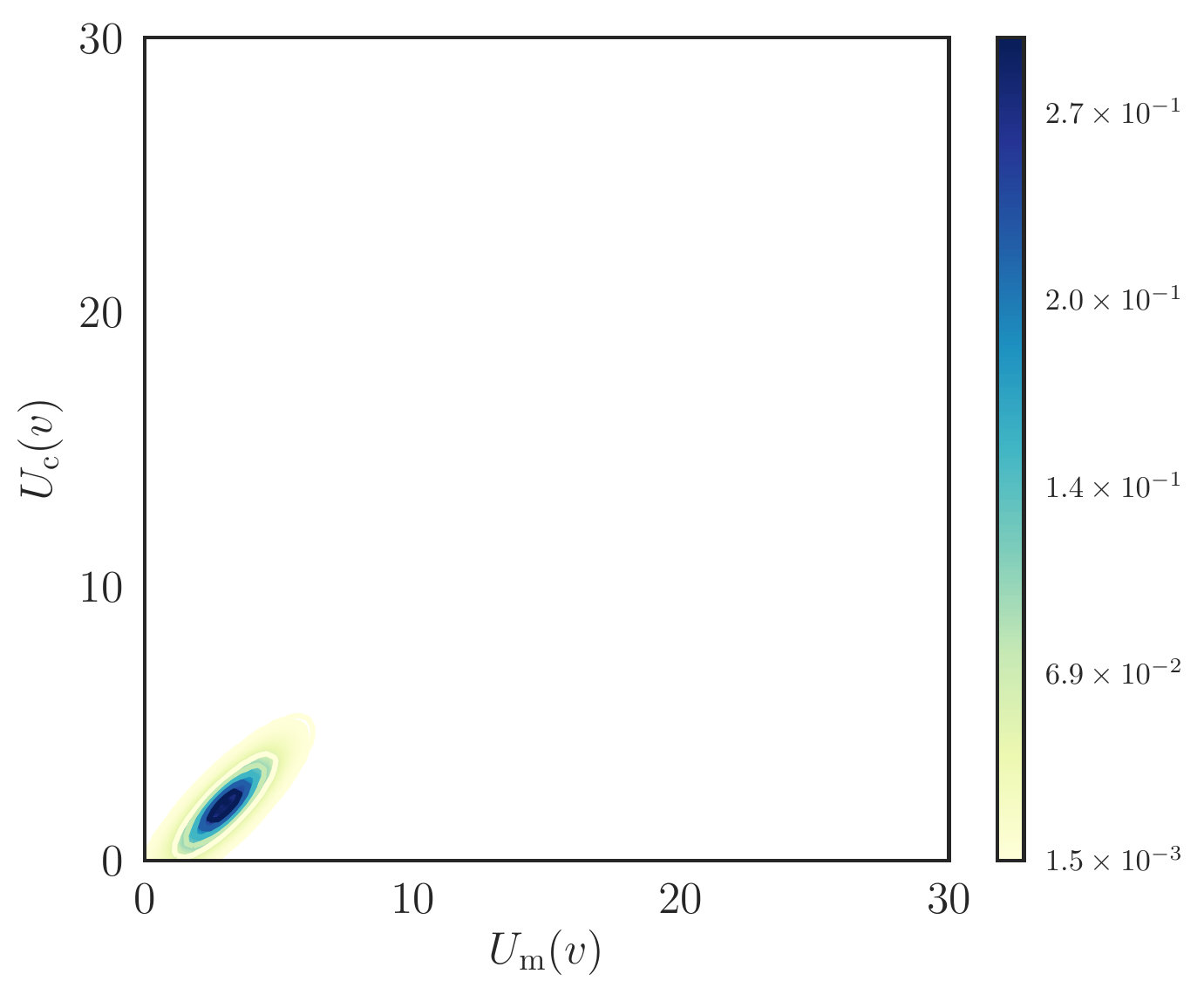}\includegraphics[width=0.48\columnwidth]{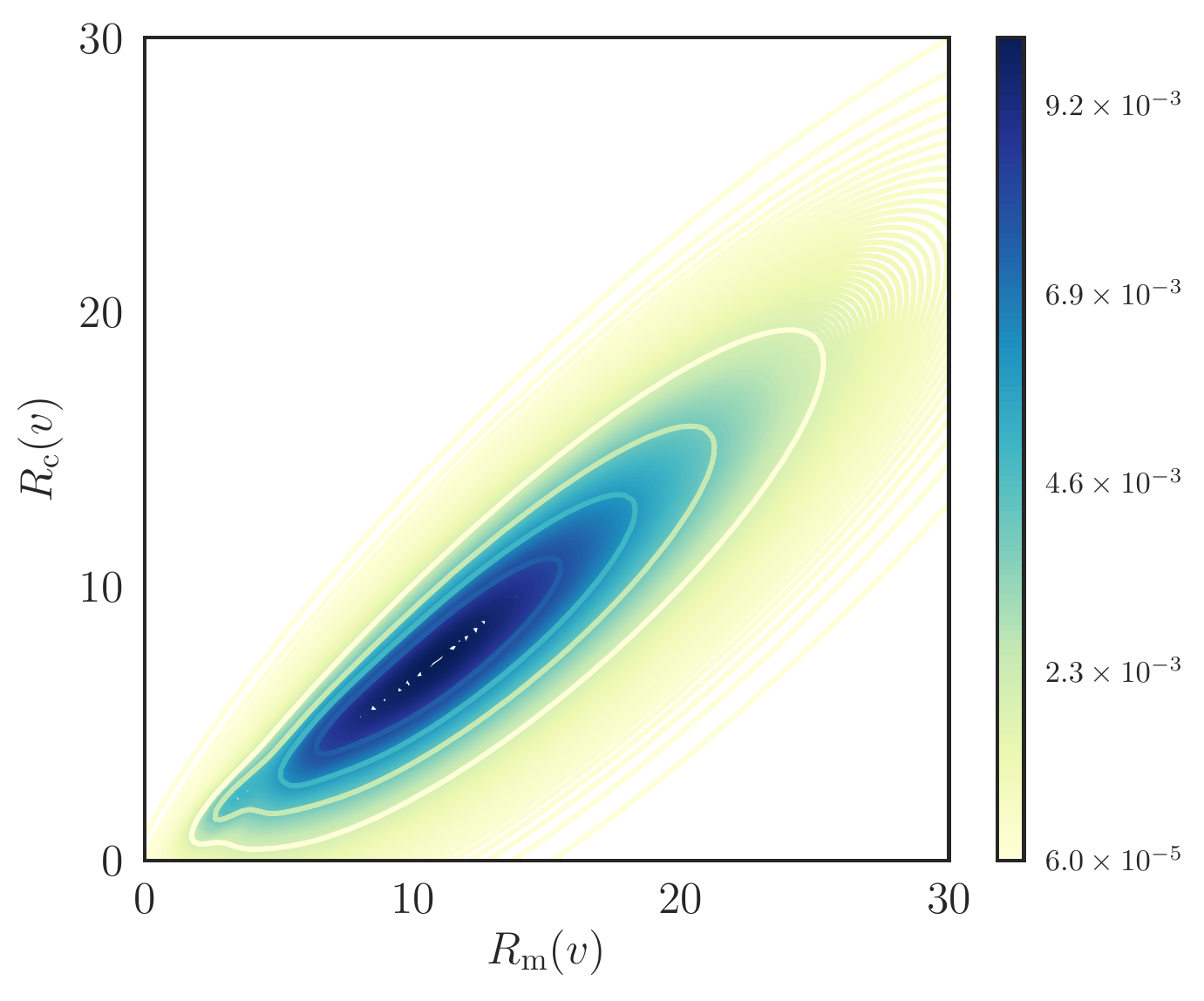}
\par\end{centering}
}
\par\end{centering}
\caption{Joint distribution $f\left(\mathbf{x};\boldsymbol{\mu}_{s},\boldsymbol{\Gamma}_{s}\right)$
(top left and bottom left) and $g\left(\mathbf{x},\boldsymbol{\mu}_{s},\boldsymbol{\Gamma}_{s}\right)$
(top right and bottom right), when $U_{s,\text{c}}\left(v\right)$
and $U_{s,\text{m}}\left(v\right)$ are (a) uncorrelated, and (b)
correlated. \label{fig:Problem_Example}}
\end{figure}

The typical joint distribution of two components of $\mathbf{U}_{s}$
and $\mathbf{R}_{s}$ is illustrated in Figure~\ref{fig:Problem_Example}.
Considering a virtual node~$v$ of a given slice~$s$, Figure~\ref{fig:Problem_Example}
represents the joint distribution $f\left(\mathbf{x};\boldsymbol{\mu}_{s},\boldsymbol{\Gamma}_{s}\right)$
of $U_{s,\text{c}}\left(v\right)$ and $U_{s,\text{m}}\left(v\right)$
and the resulting joint distribution $g\left(\mathbf{x},\boldsymbol{\mu}_{s},\boldsymbol{\Gamma}_{s}\right)$
of $R_{s,\text{c}}\left(v\right)$ and $R_{s,\text{m}}\left(v\right)$.
Here $N_{s}$ follows the binomial distribution $N_{s}\sim\mathcal{B}\left(10,0.5\right)$,
$\boldsymbol{\mu}_{s}=\left[2,3\right]^{\top}$. In Figure~\ref{fig:Prob_Ex_Uncorr},
$\boldsymbol{\Gamma}_{s}=\left[\begin{array}{cc}
1 & 0\\
0 & 1
\end{array}\right]$ is diagonal. Even if the level sets of $f\left(\mathbf{x};\boldsymbol{\mu}_{s},\boldsymbol{\Gamma}_{s}\right)$
are circles, the level sets of the resulting $g\left(\mathbf{x},\boldsymbol{\mu}_{s},\boldsymbol{\Gamma}_{s}\right)$
illustrate the correlation between $R_{s,\text{c}}\left(v\right)$
and $R_{s,\text{m}}\left(v\right)$. In Figure~\ref{fig:Prob_Ex_Un_Corr},
$\boldsymbol{\Gamma}_{s}=\left[\begin{array}{cc}
1 & 0.85\\
0.85 & 1
\end{array}\right]$ is non-diagonal, \textit{i.e}., $U_{s,\text{c}}\left(v\right)$ and
$U_{s,\text{m}}\left(v\right)$ are correlated, the correlation between
$R_{s,\text{c}}\left(v\right)$ and $R_{s,\text{m}}\left(v\right)$
increases significantly.

\subsection{Resource Consumption of Best-Effort Background Services\label{subsec:Background-Services}}

In the considered time interval, a given part of the available resources
is consumed by other best-effort background services for which no
resource provisioning has been performed. The aggregate amount of
resources consumed by these best-effort services is represented by
random variables $B_{\text{c}}\left(i\right)$ , $B_{\text{m}}\left(i\right)$
and $B_{\text{w}}\left(i\right)$, $\forall i\in\mathcal{N}$, and
$B_{\text{b}}\left(ij\right)$, $\forall ij\in\mathcal{E}$. Each
of those variables is assumed to be uncorrelated and Gaussian distributed,
$B_{n}\left(i\right)\sim\mathcal{N}\left(\mu_{\text{B},n}\left(i\right),\sigma_{\text{B},n}^{2}\left(i\right)\right)$,
$\forall i\in\mathcal{N}$, $\forall n\in\left\{ \text{c},\text{m},\text{w}\right\} $,
and $B_{n}\left(i\right)\sim\mathcal{N}\left(\mu_{\text{B},\text{b}}\left(ij\right),\sigma_{\text{B},\text{b}}^{2}\left(ij\right)\right)$,
$\forall ij\in\mathcal{E}$. Finally, denote $\mathbf{B}=\left(B_{\text{c}}\left(i\right),B_{\text{m}}\left(i\right),B_{\text{b}}\left(ij\right)\right)_{\left(i,ij\right)\in\mathcal{G}}^{\top}$
as the vector gathering all resource consumption of the background
services. $\mathbf{B}$ is distributed according to $f\left(\mathbf{x};\boldsymbol{\mu}_{\text{B}},\boldsymbol{\Gamma}_{\text{B}}\right)$,
with
\[
\boldsymbol{\mu}_{\text{B}}=\left(\mu_{\text{B},\text{c}}\left(i\right),\mu_{\text{B},\text{m}}\left(i\right),\mu_{\text{B},\text{w}}\left(i\right),\mu_{\text{B},\text{b}}\left(ij\right)\right)_{\left(i,ij\right)\in\mathcal{G}}^{\top}
\]
 and 
\[
\boldsymbol{\Gamma}_{\text{B}}=\text{diag}\left(\sigma_{\text{B},\text{c}}^{2}\left(i\right),\sigma_{\text{B},\text{m}}^{2}\left(i\right),\sigma_{\text{B},\text{w}}^{2}\left(i\right),\sigma_{\text{B},\text{b}}^{2}\left(vw\right)\right)_{\left(i,ij\right)\in\mathcal{G}}^{\top},
\]
since the elements of $\mathbf{B}$ are assumed to be uncorrelated.

\section{Optimal Slice Resource Provisioning\label{sec:Optimal-Solution}}

Consider a set of slices $\mathcal{S}$ for which infrastructure
resources have to be provisioned. To provision resource for a given
slice~$s\in\mathcal{S}$, the InP has to determine the amount of
resources each of its infrastructure nodes and links has to reserve
to satisfy the slice resource demands with a given probability. Moreover,
the InP has to preserve enough resource for background services. This
will be done by evaluating and bounding the probability that the provisioning
impacts (reduces) the resources and traffic involved by best effort
services.

\begin{figure}[tbh]
\begin{centering}
\includegraphics[width=0.55\columnwidth]{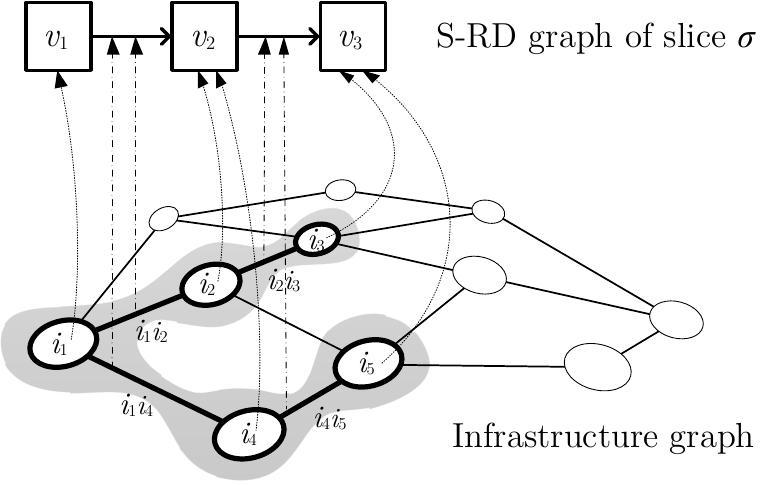}
\par\end{centering}
\caption{Provisioning of infrastructure resource to slice. In this example,
aggregate resources from the infrastructure node pair $\left(i_{1},i_{2}\right)$
is provisioned for the virtual node $v\in\mathcal{N}_{s}^{\text{R}}$
of the S-RD graph of slice $s$. Also, resources gathered from the
node pair $\left(j_{1},j_{2}\right)$ is used for $w$. Correspondingly,
two infrastructure links $\left(i_{1}j_{1}\right)$ and $\left(i_{2}j_{2}\right)$
(highlighted by the bold and dashed lines) are used to provision resource
for the virtual link $vw\in\mathcal{E}_{s}^{\text{R}}$ of the S-RD
graph of slice $s$. \label{fig:Intro_Mapping}}
\end{figure}

The slice resource provisioning is represented by a mapping between
the infrastructure graph $\mathcal{G}$ and the S-RD graph $\mathcal{G}_{s}^{\text{R}}$,
as depicted in Figure~\ref{fig:Intro_Mapping}. In this example,
the slice~$s$ consists of several linear SFCs of the same type.
The mapping has to be performed so as to minimize the provisioning
costs, while being able to satisfy the uncertain slice demands with
a high probability. The constraints that have to be satisfied by this
mapping are detailed in the following sections.

Let $\kappa_{s}\left(i,v\right)r_{s,n}(v)$ be the amount of resource
of type $n\in\left\{ \text{c},\text{m},\text{w}\right\} $ provisioned
by node~$i$ for a VNF of type~$v$, with $\kappa_{s}\left(i,v\right)\in\mathbb{N}_{0}$.
Consequently $\kappa_{s}\left(i,v\right)$ represents the number of
VNF instances of type $v\in\mathcal{N}_{s}$ that node~$i$ will
be able to host. Similarly, let $\kappa_{s}\left(ij,vw\right)r_{s,\text{b}}(vw)$
be the bandwidth provisioned by link~$ij$ to support the traffic
between virtual nodes of type~$v$ and $w$.

A solution of the provisioning problem for slice~$s$ is thus defined
by a given assignment of the variables $\boldsymbol{\kappa}_{s}=\left\{ \kappa_{s}\left(i,v\right),\kappa_{s}\left(ij,vw\right)\right\} _{\left(i,ij\right)\in\mathcal{G},\left(v,vw\right)\in\mathcal{G}_{s}^{\text{R}}}$.
This assignment has to satisfy some constraints to ensure a satisfying
behavior of the SFC and the satisfaction of the MI-SLA for slice~$s$
defined in terms of probability of satisfaction of the aggregate user
demands $\underline{p}_{s}$, see Section~\ref{subsec:Constraints}.
In addition, from the perspective of the InP, this assignment has
also to have a limited impact on the operation of background best-effort
services.

\subsection{Constraints\label{subsec:Constraints}}

Consider slice~$s$ and a given assignment of the variables $\boldsymbol{\kappa}_{s}$.
For a given node $v\in\mathcal{N}_{s}^{\text{R}}$, the probability
that enough resources are provisioned in the infrastructure network
to satisfy the resource demand $R_{s,n}\left(v\right)$ of type $n\in\left\{ \text{c},\text{m},\text{w}\right\} $
is
\begin{equation}
p_{s,n}\left(v\right)=\Pr\Big\{\sum\limits _{i}\kappa_{s}\left(i,v\right)r_{s,n}\left(v\right)\geqslant R_{s,n}\left(v\right)\Big\}.\label{eq:Cplx_Proba_Satisfy_Node}
\end{equation}
Similarly, for a given virtual link $vw\in\mathcal{E}_{s}^{\text{R}}$,
the probability that enough bandwidth is provisioned in the infrastructure
network to satisfy the demand $R_{s,\text{b}}\left(vw\right)$ is
\begin{equation}
p_{s,\text{b}}\left(vw\right)=\Pr\Big\{\sum\limits _{ij}\kappa_{s}\left(ij,vw\right)r_{s,\text{b}}\left(vw\right)\geqslant R_{s,\text{b}}\left(vw\right)\Big\}.\label{eq:Cplx_Proba_Satisfy_Link}
\end{equation}

In both cases, the assignment has to be such that, for each infrastructure
node $i\in\mathcal{N}$ and link $ij\in\mathcal{E}$, the total amount
of provisioned resources for all slices $s\in\mathcal{S}$ is less
or equal than the amount of available resources
\begin{align}
 & \sum\limits _{s,v}\kappa_{s}\left(i,v\right)r_{s,n}\left(v\right)\leqslant a_{n}\left(i\right),\label{eq:Cplx_Cons_Limit_Node}\\
 & \sum\limits _{s,vw}\kappa_{s}\left(ij,vw\right)r_{s,\text{b}}\left(vw\right)\leqslant a_{\text{b}}\left(ij\right).\label{eq:Cplx_Cons_Limit_Link}
\end{align}
The constraints \eqref{eq:Cplx_Cons_Limit_Node}-\eqref{eq:Cplx_Cons_Limit_Link}
may leave no resources for the background best-effort services. The
probability that the background best-effort services are impacted
at a node $i$ or on the link $ij$ by the provisioning for all slices~$s\in\mathcal{S}$
are, $\forall n\in\left\{ \text{c},\text{m},\text{w}\right\} $,
\begin{align}
p_{n}^{\text{im}}\left(i\right)=\Pr\Big\{\sum_{s,v}\kappa_{s}\left(i,v\right)r_{s,n}\left(v\right)\geqslant a_{n}\left(i\right)-B_{n}\left(i\right)\Big\}\label{eq:Cplx_Proba_Impact_Node}
\end{align}
and 
\begin{equation}
p_{\text{b}}^{\text{im}}\left(ij\right)=\Pr\Big\{\sum_{s,vw}\kappa_{s}\left(ij,vw\right)r_{s,\text{b}}\left(vw\right)\geqslant a_{\text{b}}\left(ij\right)-B_{\text{b}}\left(ij\right)\Big\}.\label{eq:Cplx_Proba_Impact_Link}
\end{equation}
The impact probabilities (IPs) of the provisioning for all slice $s\in\mathcal{S}$
on the nodes and links resources employed by best-effort service has
to be such that, $\forall\left(i,ij\right)\in\mathcal{G}$, $\forall n\in\left\{ \text{c},\text{m},\text{w}\right\} $
\begin{align}
 & p_{n}^{\text{im}}\left(i\right)\leqslant\overline{p}^{\text{im}},\label{eq:Cplx_Cons_Impact_Node}\\
 & p_{\text{b}}^{\text{im}}\left(ij\right)\leqslant\overline{p}^{\text{im}},\label{eq:Cplx_Cons_Impact_Link}
\end{align}
where $\overline{p}^{\text{im}}$ is the maximum tolerated impact
probability. The value of $\overline{p}^{\text{im}}$ is chosen by
the InP to provide sufficient resources for the background services
at every infrastructure nodes and links. A small value of $\overline{p}^{\text{im}}$
leads to a small impact of slice resource provisioning on background
services, but makes the provisioning problem more difficult to solve
compared to a value of $\overline{p}^{\text{im}}$ close to one.

The considered assignment has to satisfy additional constraints to
ensure that the data can be correctly carried between VNFs. For each
virtual link $vw\in\mathcal{E}_{s}^{\text{R}}$, resources on a sequence
of infrastructure links must be provisioned between \textit{each}
pair of infrastructure nodes that have provisioned resources to the
virtual nodes $v$ and $w$. One obtains a flow conservation constraint
similar to that introduced in \cite{Luu2020b}. One should have $\forall s\in\mathcal{S}$,
$\forall i\in\mathcal{N}$, $\forall vw\in\mathcal{E}_{s}$,
\begin{align}
 & \hspace{-0.2cm}\sum\limits _{j\in\mathcal{N}}\left[\kappa_{s}\left(ij,vw\right)-\kappa_{s}\left(ji,vw\right)\right]\nonumber \\
 & \hspace{-0.2cm}=\left(\frac{r_{s,\text{b}}(vw)}{{\scriptstyle \sum_{vu}}r_{s,\text{b}}(vu)}\right)\kappa_{s}\left(i,v\right)-\left(\frac{r_{s,\text{b}}(vw)}{{\scriptstyle \sum_{uw}}r_{s,\text{b}}(uw)}\right)\kappa_{s}\left(i,w\right).\label{eq:Cplx_Cons_Flow}
\end{align}

Finally, considering an assignment $\boldsymbol{\kappa}=\left\{ \boldsymbol{\kappa}_{s}\right\} _{s\in\mathcal{S}}$
which satisfies \eqref{eq:Cplx_Cons_Limit_Node}-\eqref{eq:Cplx_Cons_Flow},
the probability that this assignment is compliant with the constraints
imposed for slice~$s$ and by the infrastructure, \emph{i.e.}, the
Probability of Successful Provisioning (PSP) for slice~$s$ is
\begin{equation}
\begin{array}{clcl}
\hspace{-0.3cm}p_{s}\left(\boldsymbol{\kappa}_{s}\right)=\Pr\Big\{ & \hspace{-0.35cm}\sum\limits _{i}\kappa_{s}\left(i,v\right)r_{s,n}\left(v\right) & \hspace{-0.25cm}\geqslant & \hspace{-0.25cm}R_{s,n}\left(v\right),\forall v,n,\\
 & \hspace{-0.35cm}\sum\limits _{ij}\kappa_{s}\left(ij,vw\right)r_{s,\text{b}}\left(vw\right) & \hspace{-0.25cm}\geqslant & \hspace{-0.25cm}R_{s,\text{b}}\left(vw\right),\forall vw\Big\},
\end{array}\label{eq:Cplx_Proba_Success}
\end{equation}
and, as stated in the MI-SLA, the InP has to ensure a minimum PSP
of $\underline{p}_{s}$ for every slice~$s\in\mathcal{S}$, \emph{i.e.},
\begin{equation}
p_{s}\left(\boldsymbol{\kappa}_{s}\right)\geqslant\underline{p}_{s}.\label{eq:Cplx_Cons_Success}
\end{equation}

\subsection{Costs, Incomes, and Earnings}

Considering the perspective of the InP, this section presents the
cost, income, and earnings model for the slice resource provisioning
problem.

Consider a given slice~$s\in\mathcal{S}$ and its related assignment
of the variables $\boldsymbol{\kappa}_{s}$. Let
\begin{equation}
x_{s}\left(\boldsymbol{\kappa}_{s}\right)=\begin{cases}
1 & \text{if }p_{s}\left(\boldsymbol{\kappa}_{s}\right)\geqslant\underline{p}_{s}\\
0 & \text{else}
\end{cases}\label{eq:Indicator_xs}
\end{equation}
indicate whether the MI-SLA for slice~$s$ is satisfied.

Define $I_{s}$ as the income obtained for a slice $s$ whose MI-SLA
is satisfied. The income awarded to the InP from the MNO is then $I_{s}x_{s}\left(\boldsymbol{\kappa}_{s}\right)$.

The total provisioning cost $C_{s}$$\left(\boldsymbol{\kappa}_{s}\right)$
of a given slice~$s$ for the InP is
\begin{align}
C_{s}\left(\boldsymbol{\kappa}_{s}\right)= & \sum\limits _{i}\widetilde{\kappa}_{s}\left(i\right)c_{\textrm{f}}\left(i\right)+\sum\limits _{i,v,n}\kappa_{s}\left(i,v\right)r_{s,n}\left(v\right)c_{n}\left(i\right)\nonumber \\
+ & \sum\limits _{ij,vw}\kappa_{s}\left(ij,vw\right)r_{s,\text{b}}\left(vw\right)c_{\text{b}}\left(ij\right),\label{eq:ProvCost}
\end{align}
where
\begin{equation}
\widetilde{\kappa}_{s}\left(i\right)=\begin{cases}
1 & \text{if }\sum_{v}\kappa_{s}\left(i,v\right)>0,\\
0 & \text{otherwise.}
\end{cases}
\end{equation}
The first term of $C_{s}\left(\boldsymbol{\kappa}_{s}\right)$ represents
the fixed costs associated to the use of infrastructure nodes by slice~$s$,
whereas the second and the third terms indicate the cost of reserved
resources from infrastructure nodes and links. The variable $\widetilde{\kappa}_{s}\left(i\right)$
indicates whether the infrastructure node~$i$ is used by slice~$s$.

Finally, the total earnings $E_{s}\left(\boldsymbol{\kappa}_{s}\right)$
obtained by the InP for the successful provisioning of slice~$s$
is
\begin{equation}
E_{s}\left(\boldsymbol{\kappa}_{s}\right)=I_{s}x_{s}\left(\boldsymbol{\kappa}_{s}\right)-C_{s}\left(\boldsymbol{\kappa}_{s}\right).\label{eq:Earnings}
\end{equation}

\subsection{Nonlinear Constrained Optimization Problem}

Consider a set of slices $\mathcal{S}$, the resource provisioning
problem for all slices $s\in\mathcal{S}$, which accounts for uncertain
slice user demands and tries to limit the impact on background services,
can be formulated as
\begin{figure}[tbh]
\begin{tcolorbox}[top=0mm,bottom=0mm,left=0mm,right=0mm,colframe=black!60,title=

Problem 1: Nonlinear Constrained Optimization]\vspace{-0.2cm}
\begin{align}
 & \underset{\boldsymbol{\kappa}=\left\{ \boldsymbol{\kappa}_{s}\right\} _{s\in\mathcal{S}}}{\text{maximize}}\enskip\sum_{s\in\mathcal{S}}E_{s}\left(\boldsymbol{\kappa}_{s}\right)=\sum_{s\in\mathcal{S}}\left(I_{s}x_{s}\left(\boldsymbol{\kappa}_{s}\right)-C_{s}\left(\boldsymbol{\kappa}_{s}\right)\right),\nonumber \\
 & \text{subject to }(\ref{eq:Cplx_Cons_Limit_Node}\text{-}\ref{eq:Cplx_Cons_Limit_Link},\ref{eq:Cplx_Cons_Impact_Node}\text{-}\ref{eq:Cplx_Cons_Flow},\ref{eq:Cplx_Cons_Success}\text{-}\ref{eq:Indicator_xs}).\label{eq:Problem_Complex}
\end{align}
\end{tcolorbox}
\end{figure}

Solving Problem~1 is complex due to the need to evaluate $p_{s}\left(\boldsymbol{\kappa}_{s}\right)$
using \eqref{eq:Cplx_Proba_Success} in the verification of the constraint
\eqref{eq:Cplx_Cons_Success}. Section~\ref{subsec:MILP-Formulation}
introduces a simpler method to solve Problem~1.

\section{Reduced-Complexity Slice Resource Provisioning\label{sec:Suboptimal-Solution}}

\label{subsec:MILP-Formulation}

In this section, a parameterized MILP formulation of \eqref{eq:Problem_Complex}
is introduced. The main idea is to replace the constraints (\ref{eq:Cplx_Cons_Impact_Node},
\ref{eq:Cplx_Cons_Impact_Link}, \ref{eq:Cplx_Cons_Success}) involving
probabilities related to random variables describing the aggregate
user demands and best-effort services by linear deterministic constraints.

\subsection{Linear Inequality Constraints for the PSP \label{subsec:Linear-Constraints-PSP}}

For a given slice $s\in\mathcal{S}$ and for each $v\in\mathcal{N}_{s}$,
$vw\in\mathcal{E}_{s}$, and $n\in\left\{ \text{c},\text{m},\text{w}\right\} $,
let
\begin{align}
\overline{R}_{s,n}\left(v,\gamma_{s}\right) & =\mu_{s,n}\left(v\right)+\gamma_{s}\sigma_{s,n}\left(v\right),\label{eq:MILP_Bound_Rn}\\
\overline{R}_{s,\text{b}}\left(vw,\gamma_{s}\right) & =\mu_{\text{b}}\left(vw\right)+\gamma_{s}\sigma_{\text{b}}\left(vw\right),\label{eq:MILP_Bound_Rb}
\end{align}
be the target aggregate user demand, depending on some parameter $\gamma_{s}>0$.
For an assignment $\boldsymbol{\kappa}_{s}$ that satisfies 
\begin{align}
\sum\limits _{i}\kappa_{s}\left(i,v\right)r_{s,n}\left(v\right) & \geqslant\overline{R}_{s,n}\left(v,\gamma_{s}\right),\forall n,v,\label{eq:Relax_Satisfy_Node}\\
\sum\limits _{ij}\kappa_{s}\left(ij,vw\right)r_{s,\text{b}}\left(vw\right) & \geqslant\overline{R}_{s,\text{b}}\left(vw,\gamma_{s}\right),\forall vw,\label{eq:Relax_Satisfy_Link}
\end{align}
and (\ref{eq:Cplx_Cons_Limit_Node}, \ref{eq:Cplx_Cons_Limit_Link},
\ref{eq:Cplx_Cons_Flow}), the PSP defined in \eqref{eq:Cplx_Proba_Success}
can be evaluated as
\begin{equation}
\begin{array}{clcl}
p_{s}\left(\gamma_{s}\right)=\Pr\Big\{ & \hspace{-0.35cm}\overline{R}_{s,n}\left(v,\gamma_{s}\right) & \hspace{-0.25cm}\geqslant & \hspace{-0.25cm}R_{s,n}\left(v\right),\forall v,n,\\
 & \hspace{-0.35cm}\overline{R}_{s,\text{b}}\left(vw,\gamma_{s}\right) & \hspace{-0.25cm}\geqslant & \hspace{-0.25cm}R_{s,\text{b}}\left(vw\right),\forall vw\Big\},
\end{array}
\end{equation}
which is independent of $\boldsymbol{\kappa}_{s}.$ If $p_{s}\left(\gamma_{s}\right)\geqslant\underline{p}_{s}$,
the MI-SLA relative to the PSP is satisfied. The main difficulty is
now to determine the smallest value of $\gamma_{s}$ such that $p_{s}\left(\gamma_{s}\right)\geqslant\underline{p}_{s}$,
since the larger $\gamma_{s}$, the more difficult the satisfaction
of \eqref{eq:Relax_Satisfy_Node} and \eqref{eq:Relax_Satisfy_Link}. 

\begin{figure}[tbh]
\begin{centering}
\includegraphics[width=0.6\columnwidth]{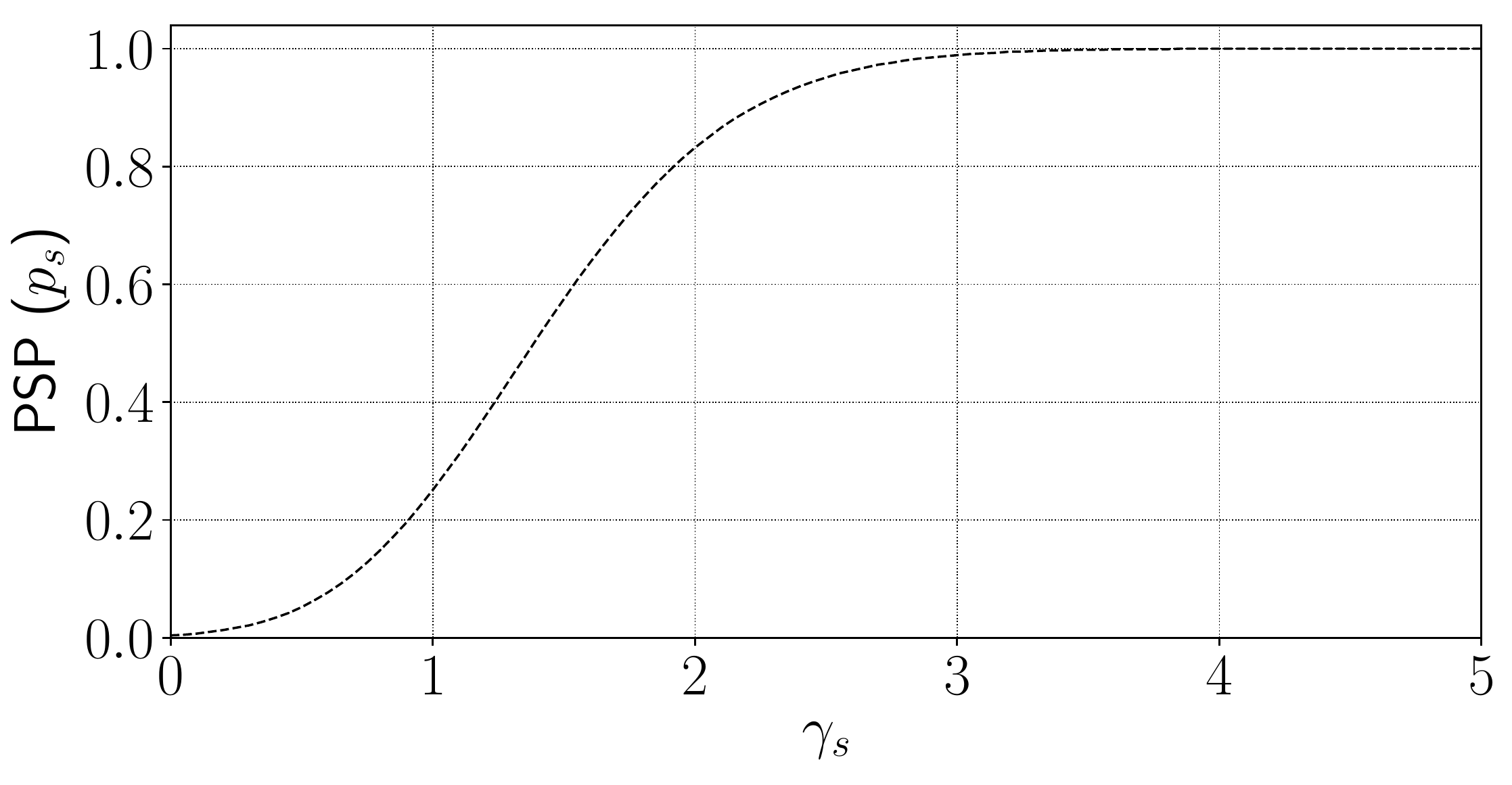}
\par\end{centering}
\caption{Evolution of $p_{s}$ as function of $\gamma_{s}$. \label{fig:Prob_ps}}
\end{figure}

Using \eqref{eq:PDF_Rs}, one has
\begin{align}
p_{s}\left(\gamma_{s}\right)=\sum\limits _{k=1}^{m}p_{k}{\displaystyle \int_{\mathcal{\overline{R}}\left(\gamma_{s}\right)}}f\left(\mathbf{x},k\boldsymbol{\mu},k^{2}\boldsymbol{\Gamma}\right)\textrm{d}\mathbf{x},\label{eq:Relax_Proba_Success}\\
\end{align}
where $\mathcal{\overline{R}}\left(\gamma_{s}\right)=\left\{ \mathbf{x}\in\mathbb{R}^{n_{\text{R}}}\,|\,\mathbf{x}\leqslant\overline{\mathbf{R}}\left(\gamma_{s}\right)\right\} $
and 
\begin{align*}
\overline{\mathbf{R}}\left(\gamma_{s}\right)=\Big( & \overline{R}_{s,\text{c}}\left(v_{1},\gamma_{s}\right),\overline{R}_{s,\text{m}}\left(v_{1},\gamma_{s}\right),\dots\\
 & \overline{R}_{s,\text{b}}\left(v_{1}v_{2},\gamma_{s}\right),\dots\Big)^{\top}
\end{align*}
of size $n_{\text{R}}$. Since the pmf of the number of users $p_{k}$,
$k=1,\dots,m$ has been assumed to be known, the value of $\gamma_{s}$
such that $p_{s}\left(\gamma_{s}\right)=\underline{p}_{s}$ may be
obtained by dichotomy search. The multidimensional integral in \eqref{eq:Relax_Proba_Success}
can be evaluated using a quasi-Monte Carlo integration algorithm presented
in \cite{Genz2004}. An example of the evolution of $p_{s}\left(\gamma_{s}\right)$
as function of $\gamma_{s}$ for a given slice~$s$ of Type 1 is
depicted in Figure~\ref{fig:Prob_ps}, using the simulation setting
described in Section~\ref{subsec:SetUp}.

\subsection{Linear Inequality Constraints for the IP\label{subsec:Linear-Constraints-IP}}

For each $i\in\mathcal{N}$, $ij\in\mathcal{E}$, and $n\in\left\{ \text{c},\text{m},\text{w}\right\} $,
consider the following target level of background service demands
\begin{align}
\overline{B}_{n}\left(i,\gamma_{\text{B}}\right) & =\mu_{\text{B},n}\left(i\right)+\gamma_{\text{B}}\sigma_{\text{B},n}\left(i\right),\label{eq:MILP_Bound_Bn}\\
\overline{B}_{\text{b}}\left(ij,\gamma_{\text{B}}\right) & =\mu_{\text{B},\text{b}}\left(ij\right)+\gamma_{\text{B}}\sigma_{\text{B},\text{b}}\left(ij\right),\label{eq:MILP_Bound_Bb}
\end{align}
where $\gamma_{\text{B}}>0$ is some tuning parameter. For an assignment
$\boldsymbol{\kappa}=\left\{ \boldsymbol{\kappa}_{s}\right\} _{s\in\mathcal{S}}$
that satisfies
\begin{align}
\sum\limits _{s,v}\kappa_{s}\left(i,v\right)r_{s,n}\left(v\right) & \leqslant a_{n}\left(i\right)-\overline{B}_{n}\left(i,\gamma_{\text{B}}\right),\forall n,i,\label{eq:Relax_Limit_Node}\\
\sum\limits _{s,vw}\kappa_{s}\left(ij,vw\right)r_{s,\text{b}}\left(vw\right) & \leqslant a_{\text{b}}\left(ij\right)-\overline{B}_{\text{b}}\left(ij,\gamma_{\text{B}}\right),\forall ij,\label{eq:Relax_Limit_Link}
\end{align}
and (\ref{eq:Cplx_Cons_Limit_Node}, \ref{eq:Cplx_Cons_Limit_Link},
\ref{eq:Cplx_Cons_Flow}), the IP defined in \eqref{eq:Cplx_Proba_Impact_Node}
can be evaluated as follows
\begin{align}
p_{n}^{\text{im}}\left(i\right) & =\Pr\Big\{ B_{n}\left(i\right)\geqslant\overline{B}_{n}\left(i,\gamma_{\text{B}}\right)\Big\}\nonumber \\
 & =\int_{\overline{B}_{n}\left(i,\gamma_{\text{B}}\right)}^{+\infty}\hspace{-0.2cm}f\left(x;\mu_{\text{B},n}\left(i\right),\sigma_{\text{B},n}^{2}\left(i\right)\right)\textrm{d}x\nonumber \\
 & =1-\int_{-\infty}^{\overline{B}_{n}\left(i,\gamma_{\text{B}}\right)}\hspace{-0.2cm}f\left(x;\mu_{\text{B},n}\left(i\right),\sigma_{\text{B},n}^{2}\left(i\right)\right)\textrm{d}x\nonumber \\
 & =1-\Phi\left(\gamma_{\text{B}}\right),\label{eq:Relax_Proba_Impact_Node}
\end{align}
where $\Phi$ is the cumulative distribution function (CDF) of the
zero-mean, unit-variance normal distribution. Similarly, the IP defined
in \eqref{eq:Cplx_Proba_Impact_Link} can also be evaluated as
\begin{align}
p_{s,\text{b}}^{\text{im}}\left(ij\right) & =\Pr\Big\{ B_{\text{b}}\left(ij\right)\geqslant\overline{B}_{\text{b}}\left(ij,\gamma_{\text{B}}\right)\Big\}\nonumber \\
 & =1-\Phi\left(\gamma_{\text{B}}\right).\label{eq:Relax_Proba_Impact_Link}
\end{align}
Both \eqref{eq:Relax_Proba_Impact_Node} and \eqref{eq:Relax_Proba_Impact_Link}
are independent of $\boldsymbol{\kappa}_{s}$, $\forall s\in\mathcal{S}$.
To satisfy the impact constraints imposed by (\ref{eq:Cplx_Proba_Impact_Node},
\ref{eq:Cplx_Proba_Impact_Link}), $\gamma_{\text{B}}$ has to be
chosen such that
\begin{align}
1-\Phi\left(\gamma_{\text{B}}\right)\leqslant\overline{p}^{\text{im}} & \Leftrightarrow\gamma_{\text{B}}\geqslant\Phi^{-1}\left(1-\overline{p}^{\text{im}}\right).\label{eq:Relax_Gamma_B}
\end{align}
Since the larger $\gamma_{\text{B}}$, the more difficult the satisfaction
of \eqref{eq:Relax_Limit_Node} and \eqref{eq:Relax_Limit_Link},
the optimal $\gamma_{\text{B}}$ would be $\gamma_{\text{B}}=\Phi^{-1}\left(1-\overline{p}^{\text{im}}\right)$.

\subsection{MILP Formulation for Multiple Slice Provisioning\label{subsec:JointMILP}}

Considering the linear inequality constraints introduced in Sections~\ref{subsec:Linear-Constraints-PSP}
and \ref{subsec:Linear-Constraints-IP} instead of the inequality
constraints involving probabilities in Problem~1, one may introduce
the following relaxed parameterized formulation of Problem~1.

\begin{figure}[tbh]
\begin{tcolorbox}[top=0mm,bottom=0mm,left=0mm,right=0mm,colframe=black!60,title=

Problem 2: MILP for Multiple Slice Resource Provisioning ]\vspace{-0.2cm}
\begin{align}
 & \underset{\left\{ \boldsymbol{d},\boldsymbol{\kappa}\right\} =\left\{ d_{s},\boldsymbol{\kappa}_{s}\right\} _{s\in\mathcal{S}}}{\text{maximize}}\enskip\sum_{s\in\mathcal{S}}\left(I_{s}d_{s}-C_{s}\left(\boldsymbol{\kappa}_{s}\right)\right),\label{eq:Problem_MILP}\\
 & \text{subject to }\eqref{eq:Cplx_Cons_Flow}\text{ and}\nonumber \\
 & \sum\limits _{i}\kappa_{s}\left(i,v\right)r_{s,n}\left(v\right)\geqslant\overline{R}_{s,n}\left(v,\gamma_{s}\right)d_{s},\forall s,n,v,\label{eq:MILP_Cons_Satisfy_Node}\\
 & \sum\limits _{ij}\kappa_{s}\left(ij,vw\right)r_{s,\text{b}}\left(vw\right)\geqslant\overline{R}_{s,\text{b}}\left(vw,\gamma_{s}\right)d_{s},\forall s,vw,\label{eq:MILP_Cons_Satisfy_Link}\\
 & \sum\limits _{s,v}\kappa_{s}\left(i,v\right)r_{s,n}\left(v\right)\leqslant a_{n}\left(i\right)-\overline{B}_{n}\left(i,\gamma_{\text{B}}\right),\forall n,i,\label{eq:MILP_Cons_Limit_Node}\\
 & \sum\limits _{s,vw}\kappa_{s}\left(ij,vw\right)r_{s,\text{b}}\left(vw\right)\leqslant a_{\text{b}}\left(ij\right)-\overline{B}_{\text{b}}\left(ij,\gamma_{\text{B}}\right),\forall ij.\label{eq:MILP_Cons_Limit_Link}
\end{align}
\end{tcolorbox}
\end{figure}

Problem~2 is now an MILP. The binary variables $d_{s}$, $s\in\mathcal{S}$
indicate whether resources are actually provisioned for slice $s$.
When $d_{s}=0$, the minimization of the provisioning cost $C_{s}\left(\boldsymbol{\kappa}_{s}\right)$
imposed by \eqref{eq:Problem_MILP} will enforce $\kappa_{s}=0$ in
\eqref{eq:MILP_Cons_Satisfy_Node} and \eqref{eq:MILP_Cons_Satisfy_Link}.
Remind that $\gamma_{s}$ and $\gamma_{\text{B}}$ are evaluated by
dichotomy search, as discussed in Sections~\ref{subsec:Linear-Constraints-PSP}
and \ref{subsec:Linear-Constraints-IP}, before solving Problem~2.

\subsection{MILP Formulation for Slice-by-Slice Provisioning\label{subsec:SequentialMILP}}

The number of variables involved in the solution of Problem~2 introduced
in Section~\ref{subsec:JointMILP} may be relatively large when several
slices have to be considered jointly. This section introduces a reduced-complexity
formulation where provisioning is performed slice-by-slice. 

Consider the set of $n_{\text{s}}$ slices $\mathcal{S}=\left\{ s_{1},\dots,s_{n_{\text{s}}}\right\} $
for which resources have to be provisioned. Assume that the the slice-by-slice
resource provisioning has been performed up to slice $s_{\ell-1}$,
$1\leqslant\ell-1<n_{\text{s}}$. A successful provisioning is indicated
by $d_{s}=1$, whereas $d_{s}=0$ indicates that resources cannot
be provisioned for slice~$s$, due, \emph{e.g.}, to the non-satisfaction
of the PSP or IP constraints, or to the lack of infrastructure resources.
The corresponding assignment is represented by $\boldsymbol{\kappa}_{s}$,
$s\in\left\{ s_{1},\dots,s_{\ell-1}\right\} $.

Slice~$s_{\ell}$ is now considered. In the provisioning for slice~$s_{\ell}$,
one has simply to account for the amount of infrastructure resources
left after the provisioning of all slices $s\in\left\{ s_{1},\dots,s_{\ell-1}\right\} $.
Consequently, only \eqref{eq:MILP_Cons_Limit_Node} and \eqref{eq:MILP_Cons_Limit_Link}
have to be updated to get the following new MILP formultaion for slice-by-slice
resource provisioning.

\begin{figure}[tbh]
\begin{tcolorbox}[top=0mm,bottom=0mm,left=0mm,right=0mm,colframe=black!60,title=

Problem 3: MILP for Slice-by-Slice Resource Provisioning ]\vspace{-0.2cm}
\begin{align}
 & \underset{d_{s_{\ell}},\boldsymbol{\kappa}_{s_{\ell}}}{\text{maximize}}\enskip I_{s_{\ell}}d_{s_{\ell}}-C_{s_{\ell}}\left(\boldsymbol{\kappa}_{s_{\ell}}\right),\label{eq:Problem_MILP-1}\\
 & \text{subject to }\eqref{eq:Cplx_Cons_Flow}\text{ and}\nonumber \\
 & \sum\limits _{i}\kappa_{s_{\ell}}\left(i,v\right)r_{s,n}\left(v\right)\geqslant\overline{R}_{s_{\ell},n}\left(v,\gamma_{s_{\ell}}\right)d_{s_{\ell}},\forall n,v,\label{eq:MILP_Cons_Satisfy_Node-1}\\
 & \sum\limits _{ij}\kappa_{s_{\ell}}\left(ij,vw\right)r_{s,\text{b}}\left(vw\right)\geqslant\overline{R}_{s_{\ell},\text{b}}\left(vw,\gamma_{s_{\ell}}\right)d_{s_{\ell}},\forall vw,\label{eq:MILP_Cons_Satisfy_Link-1}\\
 & \sum\limits _{v}\kappa_{s_{\ell}}\left(i,v\right)r_{s,n}\left(v\right)\leqslant a_{n}\left(i\right)-\overline{B}_{n}\left(i,\gamma_{\text{B}}\right)\\
 & \hspace{2.8cm}-\hspace{-0.7cm}\sum_{s\in\left\{ s_{1},\dots,s_{\ell-1}\right\} }\hspace{-0.7cm}\kappa_{s}\left(i,v\right)r_{s,n}\left(v\right)d_{s},\forall n,i,\label{eq:MILP_Cons_Limit_Node-1}\\
 & \sum\limits _{vw}\kappa_{s_{\ell}}\left(ij,vw\right)r_{s,\text{b}}\left(vw\right)\leqslant a_{\text{b}}\left(ij\right)-\overline{B}_{\text{b}}\left(ij,\gamma_{\text{B}}\right)\\
 & \hspace{2.8cm}-\hspace{-0.7cm}\sum_{s\in\left\{ s_{1},\dots,s_{\ell-1}\right\} }\hspace{-0.7cm}\kappa_{s}\left(ij,vw\right)r_{s,\text{b}}\left(vw\right)d_{s},\forall ij.\label{eq:MILP_Cons_Limit_Link-1}
\end{align}
\end{tcolorbox}
\end{figure}

The order in which the provisioning is performed is important. One
may choose to provision the slices by decreasing income $I_{s}$.
An other possibility is to perform a greedy search, starting with
the slice~$s^{1}\in\mathcal{S}$ for which $I_{s}d_{s}-C_{s}\left(\boldsymbol{\kappa}_{s}\right)$
is maximized, when deployed alone. Then, assuming that resources have
been provisioned for $s^{1}$, one may search $s^{2}\in\mathcal{S}\setminus\left\{ s^{1}\right\} $
maximizing $I_{s}d_{s}-C_{s}\left(\boldsymbol{\kappa}_{s}\right)$
with the remaining infrastructure resources, and so forth. 

\section{Evaluation\label{sec:Evaluation}}

In this section, one evaluates via simulations the performance of
the provisioning algorithms described in Section~\ref{sec:Suboptimal-Solution}.
Four variants based on the suboptimal method are compared. The joint
($\mathtt{JP\text{-}B}$) and sequential ($\mathtt{SP\text{-}B}$)
slice resource provisioning approaches account for the impact of provisioning
on background services, whereas the conventional joint ($\mathtt{JP}$)
and sequential ($\mathtt{SP}$) provisioning methods do not take those
services into account. This is obtained by setting $\overline{B}_{n}\left(i,\gamma_{\text{B}}\right)=0,\forall n,i$
and $\overline{B}_{\text{b}}\left(ij,\gamma_{\text{B}}\right)=0,\forall ij$
in Problems~2 and~3.

The simulation setup is described in Section~\ref{subsec:SetUp}.
All simulations are performed with the CPLEX MILP solver interfaced
with MATLAB.

\subsection{Simulation Conditions\label{subsec:SetUp}}

\subsubsection{Infrastructure Topology}

\begin{figure}[tbh]
\begin{centering}
\includegraphics[width=0.6\columnwidth]{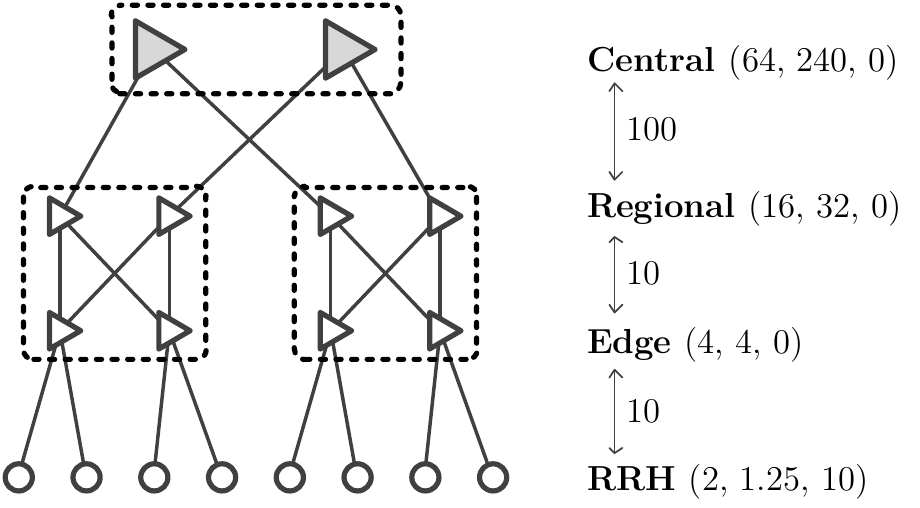}
\par\end{centering}
\caption{Description of a $k$-ary fat-tree infrastructure network with $k=2$;
Nodes provide a given amount of computing $a_{\textrm{c}}$, memory
$a_{\textrm{m}}$, and wireless $a_{\textrm{w}}$ resources expressed
in number of used CPUs, Gbytes, and Gbps; Links are able to transmit
data at a rate $a_{\text{b}}$ expressed in Gbps.\label{fig:Description_Infra}}
\end{figure}

The infrastructure network is generated from a $k$-ary fat tree topology,
as in \cite{Riggio2016,Bouten2017}. A typical fat-tree topology is
depicted in Figure~\ref{fig:Description_Infra} when $k=2$. The
leaf nodes represent the Remote Radio Heads (RRHs). The other nodes
represent the edge, regional, and central data centers. Infrastructure
nodes and links provide a given amount of computing, storage, and
possibly wireless resources $\left(a_{\textrm{c}},a_{\textrm{m}},a_{\textrm{w}}\right)$,
expressed in number of CPUs, Gbytes, and Gbps, depending on the layer
they are located. The cost of using each resource of the infrastructure
network is $c_{n}\left(i\right)=1$, $\forall n\in\left\{ \text{c},\text{m},\text{w}\right\} $,
$c_{\text{f}}\left(i\right)=65$, $60$, $55$, $50$ for respectively
central, regional, edge, RRH nodes, and $c_{\text{b}}\left(ij\right)=1$,
$\forall ij\in\mathcal{E}$.

\subsubsection{Background Services\label{subsec:Eva_Background-Services}}

At each infrastructure node $i\in\mathcal{N}$ and link~$ij\in\mathcal{E}$,
the resources consumed by best-effort background services follow a
normal distribution with mean and standard deviation equal to respectively
$20\,\%$ and $5\,\%$ percent of the available resource at that node
and link, \textit{i.e}., $\mu_{\text{B},n}\left(i\right)=0.2a_{n}\left(i\right)$,
$\sigma_{\text{B},n}\left(i\right)=0.05a_{n}\left(i\right)$, $\forall i\in\mathcal{N}$,
$\forall n\in\left\{ \text{c},\text{m,\text{w}}\right\} $, and $\mu_{\text{B},\text{b}}\left(ij\right)=0.2a_{\text{b}}\left(ij\right)$,
$\sigma_{\text{B},\text{b}}\left(ij\right)=0.05a_{\text{b}}\left(ij\right)$,
$\forall ij\in\mathcal{E}$.

\subsubsection{Slice Resource Demand (S-RD)\label{subsec:Eva_Slice-Resource-Demand}}

Three types of slices are considered.
\begin{itemize}
\item Slices of type~1 aim to provide an HD video streaming service at
average rate of$4$~Mbps for VIP users, \emph{e.g.}, in a stadium.
The number of users follows a binomial distribution $\mathcal{B}\left(300,0.9\right)$;
\item Slices of type~2 are dedicated to provide an SD video streaming service
at average rate of $2$~Mbps. The number of users follows a binomial
distribution $\mathcal{B}\left(1000,0.8\right)$;
\item Slices of type~3 aim to provide a video surveillance and traffic
monitoring service at average rate of $1$~Mbps for $100$ cameras,
\emph{e.g.}, installed along a highway.
\end{itemize}
The first two slice types address a video streaming service, and thus
have the same function architecture with $3$ virtual functions: a
virtual Video Optimization Controller (vVOC), a virtual Gateway (vGW),
and a virtual Base Band Unit (vBBU). The third slice type consists
of five virtual functions: a vBBU, a vGW, a virtual Traffic Monitor
(vTM), a vVOC, and a virtual Intrusion Detection Prevention System
(vIDPS).

As detailed in Section~\ref{subsec:Graphs-of-Resource-Demands},
the resource requirements for the various SFCs that will have to be
deployed within a slice are aggregated within an S-RD graph that mimics
the SFC-RD graph. S-RD nodes and links are characterized by the aggregated
resource needed to support the targeted number of users. Details of
each resource type as well as the associated U-RD, SFC-RD, and S-RD
graph are given in Table~\ref{tab:SRD}. Numerical values in Table~\ref{tab:SRD}
have been adapted from \cite{Savi2017}.

\subsection{Results\label{subsec:Eva-Results}}

This section illustrates the performance of the four resource provisioning
variants ($\mathtt{JP}$, $\mathtt{SP}$, $\mathtt{JP\text{-}B}$,
and $\mathtt{SP\text{-}B}$), in terms of: utilization of infrastructure
nodes and links, maximal probability of impact $p^{\text{im}}$ on
the background services at every infrastructure node and link, provisioning
cost, total earnings of the InP, and the number of impacted nodes
and links, \emph{i.e.}, the number of nodes~$i\in\mathcal{N}$ such
that $\exists n\in\left\{ \text{c},\text{m},\text{w}\right\} $ $p_{n}^{\text{im}}\left(i\right)>\overline{p}^{\text{im}}$
and links~$ij\in\mathcal{E}$ such that $p_{\text{b}}^{\text{im}}\left(ij\right)>\overline{p}^{\text{im}}$.

\subsubsection{Provisioning of a Single Slice\label{subsec:Eva_Single}}

Table~\ref{tab:Eva-Single-SNB-vs-SN} shows the performance of two
variants $\mathtt{SP\text{-}B}$ and $\mathtt{SP}$ for the provisioning
of a single slice of Type~1, where $\underline{p}_{s}=0.99$ and
$\overline{p}^{\text{im}}=0.1$. It is observed that the $\mathtt{SP}$
variant, which does not account for impact on background services,
has a lower link usage and provisioning cost, and yields a higher
earning for the InP than that of the $\mathtt{SP\text{-}B}$ variant.
Nevertheless, as expected, the $\mathtt{SP}$ variant has a higher
impact on background services, with maximal impact probability of
$0.58$ exceeding the maximum tolerated impact probability $\overline{p}^{\text{im}}$
at one infrastructure node, as summarized in Table~\ref{tab:Eva-Single-SNB-vs-SN}.

\begin{table}[H]
\caption{Performance of $\mathtt{SP\text{-}B}$ and $\mathtt{SP}$ on Single
Slice Provisioning\label{tab:Eva-Single-SNB-vs-SN}}

\centering  
\begin{tabular}{p{0.1\columnwidth}p{0.1\columnwidth}p{0.1\columnwidth}}  
\toprule  

\multicolumn{1}{c}{ Criteria }&\multicolumn{1}{c}{ $\mathtt{SP\text{-}B}$
}&\multicolumn{1}{c}{ $\mathtt{SP}$ }\\

\cmidrule[0.4pt](lr{0.12em}){1-1}%
\cmidrule[0.4pt](lr{0.12em}){2-2}%
\cmidrule[0.4pt](lr{0.12em}){3-3}%

\multicolumn{1}{c}{ Node usage }&\multicolumn{1}{c}{ $33\%$ }&\multicolumn{1}{c}{
$33\%$ }\\

\multicolumn{1}{c}{ Link usage }&\multicolumn{1}{c}{ $28\%$ }&\multicolumn{1}{c}{
$25\%$ }\\

\multicolumn{1}{c}{ Maximal $p^{\text{im}}$ }&\multicolumn{1}{c}{
$1.26\text{e-}4$ }&\multicolumn{1}{c}{ $0.58$ }\\

\multicolumn{1}{c}{ Provisioning cost }&\multicolumn{1}{c}{ $332$
}&\multicolumn{1}{c}{ $326$ }\\

\multicolumn{1}{c}{ Total earnings }&\multicolumn{1}{c}{ $568$
}&\multicolumn{1}{c}{ $574$ }\\

\multicolumn{1}{c}{ \#impacted nodes }&\multicolumn{1}{c}{ $0$
}& \multicolumn{1}{c}{ $1$ }\\

\multicolumn{1}{c}{ \#impacted links }&\multicolumn{1}{c}{ $0$
}& \multicolumn{1}{c}{ $0$ }\\

\bottomrule
\end{tabular} 
\end{table}

\begin{figure}[H]
\begin{centering}
\subfloat[Provisioning cost\label{fig:Eva_Single_SB_Cost}]{\begin{centering}
\includegraphics[width=0.23\textwidth]{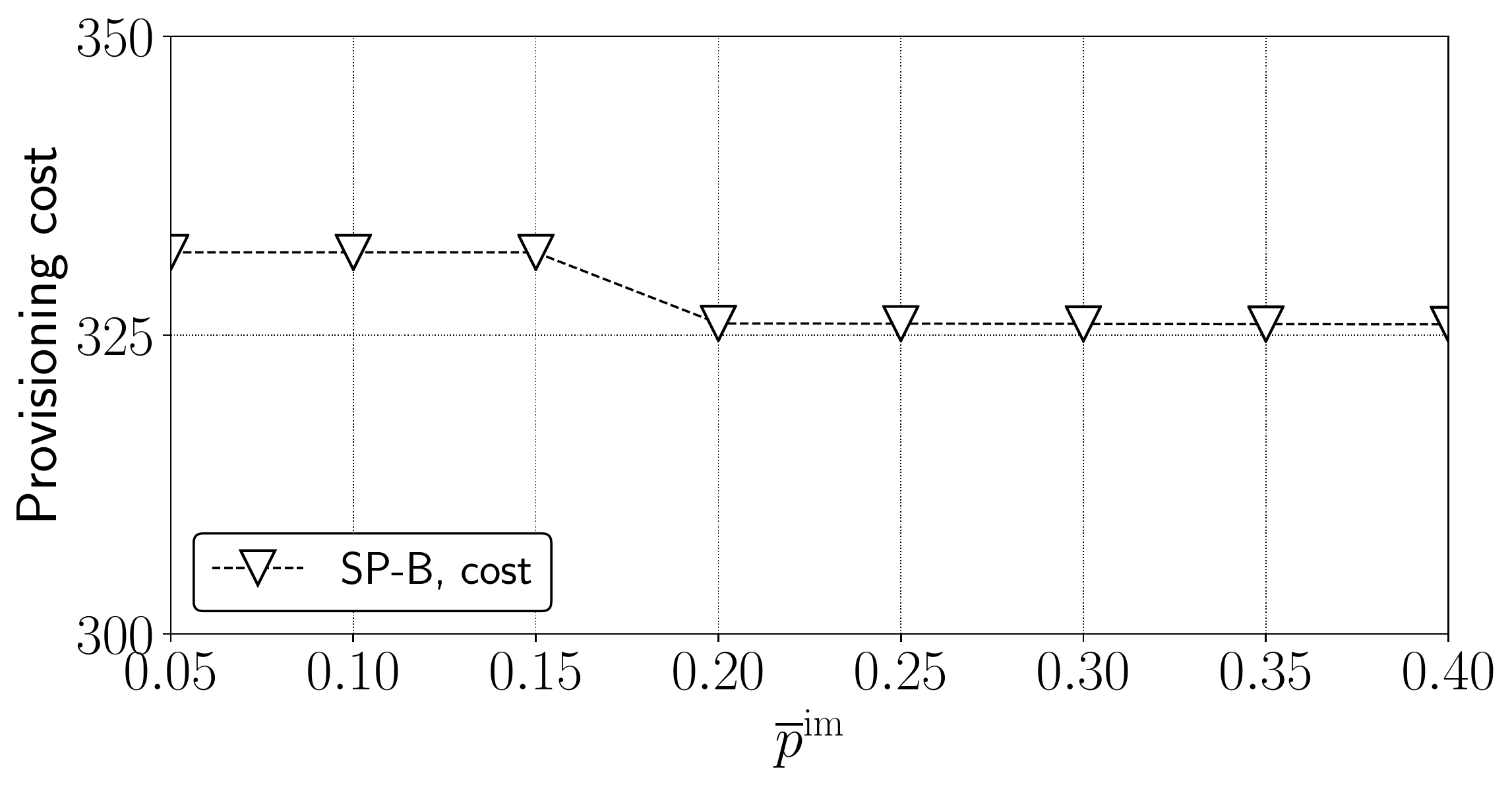}
\par\end{centering}
}\subfloat[Total earnings\label{fig:Eva_Single_SB_Earn}]{\begin{centering}
\includegraphics[width=0.23\textwidth]{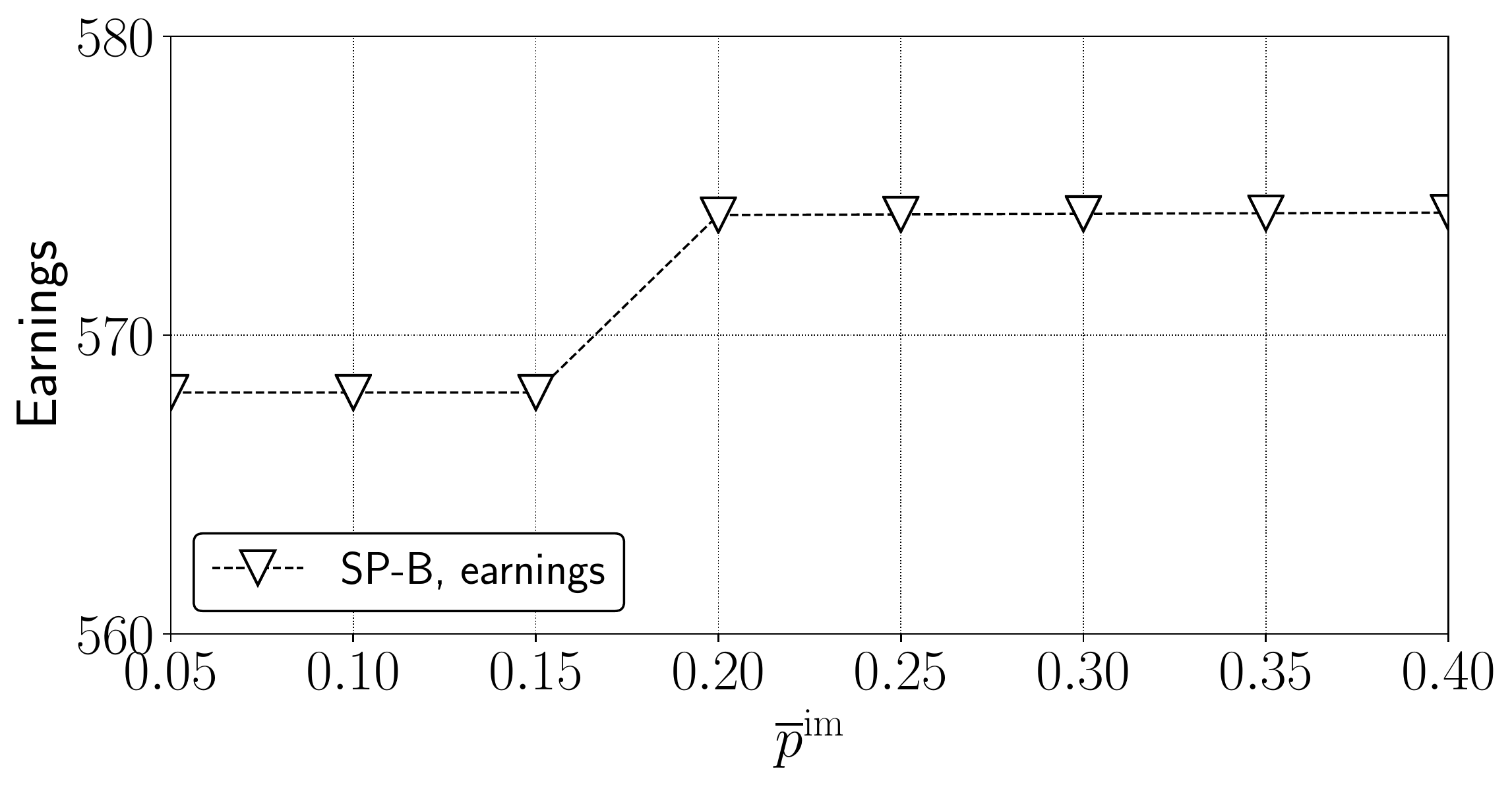}
\par\end{centering}
}
\par\end{centering}
\begin{centering}
\subfloat[Node and link usage\label{fig:Eva_Single_SB_PerNodeLink}]{\begin{centering}
\includegraphics[width=0.23\textwidth]{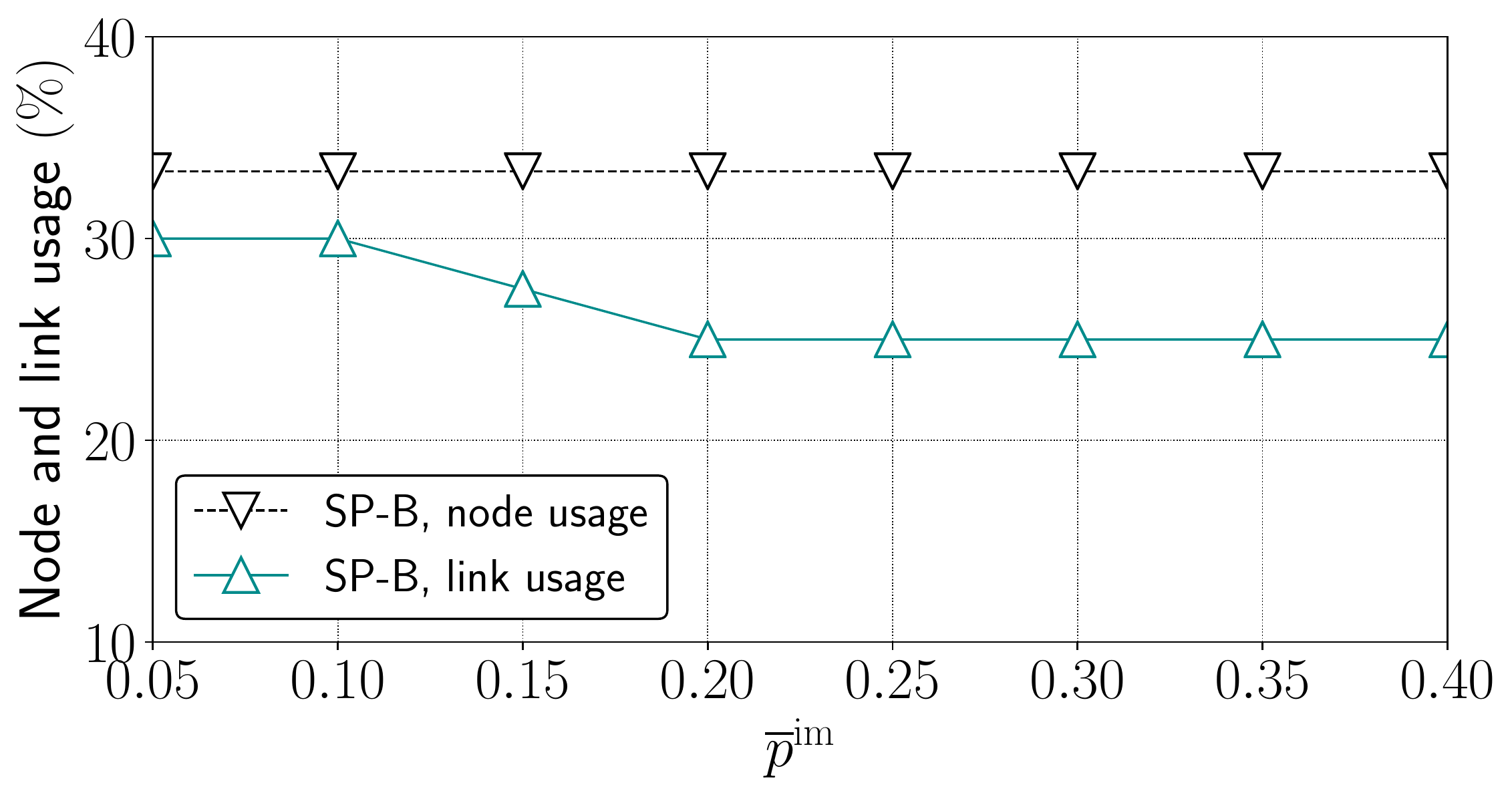}
\par\end{centering}
}\subfloat[Evolution of $p^{\text{im}}$ as a function of $\overline{p}^{\text{im}}$\label{fig:Eva_Single_SB_Impact}]{\begin{centering}
\includegraphics[width=0.23\textwidth]{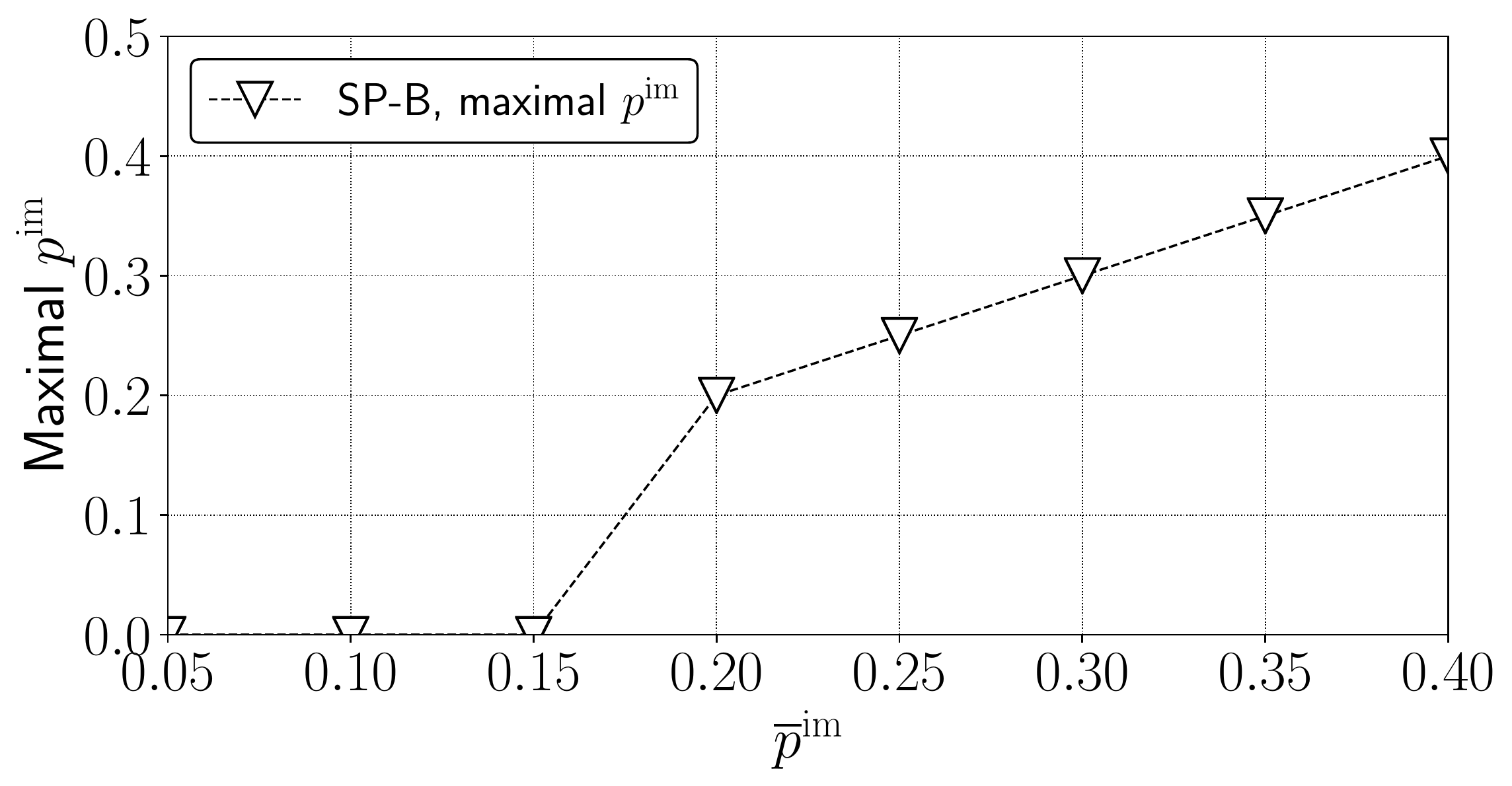}
\par\end{centering}
}
\par\end{centering}
\caption{Performance of the $\mathtt{SP\text{-}B}$ approach on single slice
provisioning problem with different values of $\overline{p}^{\text{im}}$,
in terms of (a) provisioning cost, (b) total earnings, (c) node and
link utilization, and (d) maximal impact probability $p^{\text{im}}$.
\label{fig:Eva_Single_SB}}
\end{figure}

The way $\overline{p}^{\text{im}}$ affects the performance of $\mathtt{SP\text{-}B}$
is shown in Figures~\ref{fig:Eva_Single_SB_Cost}-\ref{fig:Eva_Single_SB_Impact},
when $\underline{p}_{s}=0.99$ and $\overline{p}^{\text{im}}$ ranges
from $0.05$ to $0.4$. One observes that, the higher $\overline{p}^{\text{im}}$,
the lower the provisioning cost and the higher earnings for the InP.
This is due to the fact that, with higher $\overline{p}^{\text{im}}$,
it is easier to provision slices with limited resources. This can
be observed in the decrease of link usage in Figure~\ref{fig:Eva_Single_SB_PerNodeLink}.
On the other hand, the impact probability $p^{\text{im}}$ is always
kept under the threshold $\overline{p}^{\text{im}}$ imposed by the
InP, as shown in Figure~\ref{fig:Eva_Single_SB_Impact}.

\subsubsection{Provisioning Several Slices of the Same Type\label{subsec:Eva_Multi_1Type}}

\begin{figure}[tbh]
\begin{centering}
\subfloat[Acceptance rate\label{fig:Eva_Multi1_Accept_Rate}]{\begin{centering}
\includegraphics[width=0.23\textwidth]{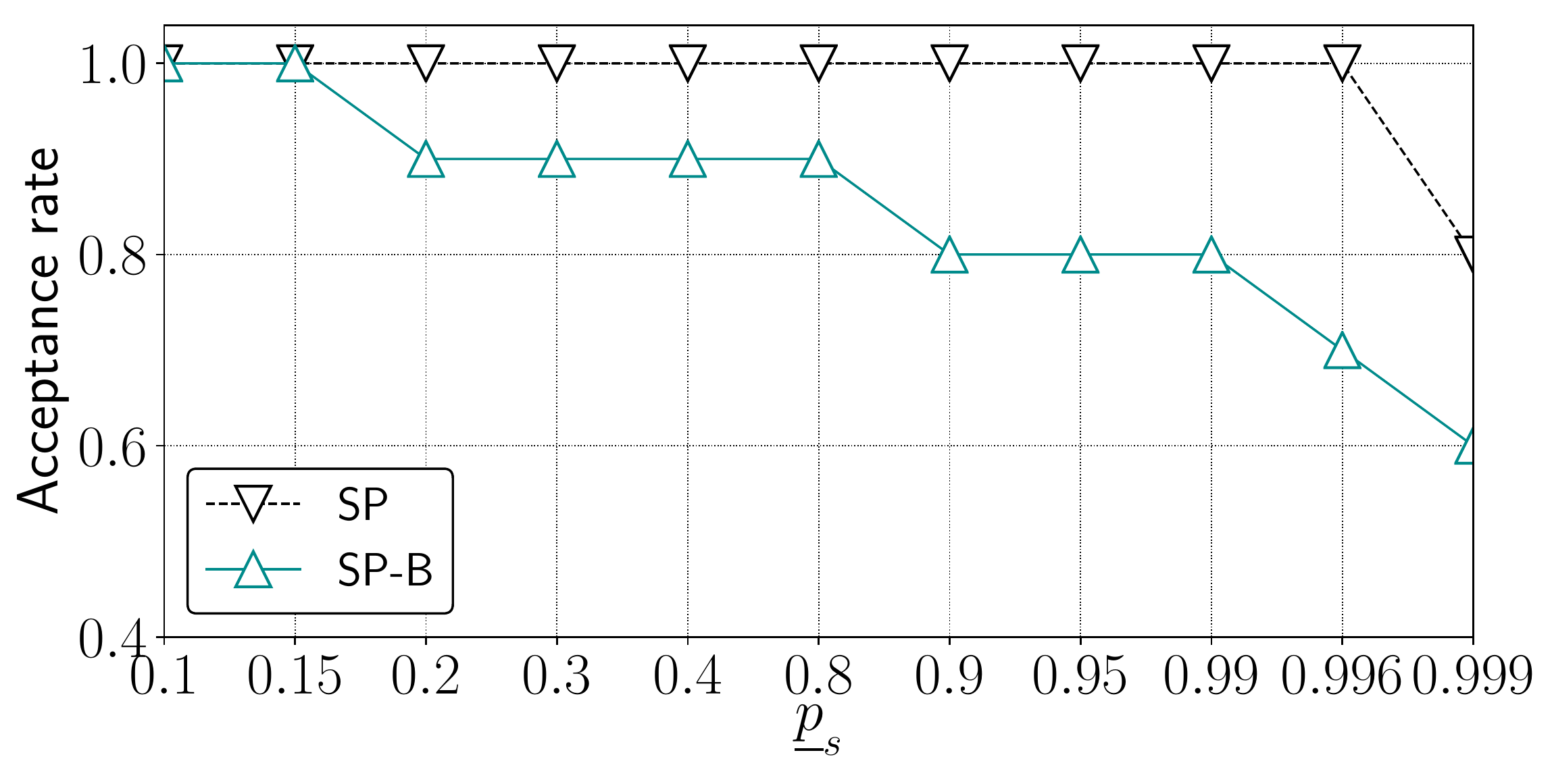}
\par\end{centering}
}\subfloat[Total earnings\label{fig:Eva_Multi1_Earn}]{\begin{centering}
\includegraphics[width=0.23\textwidth]{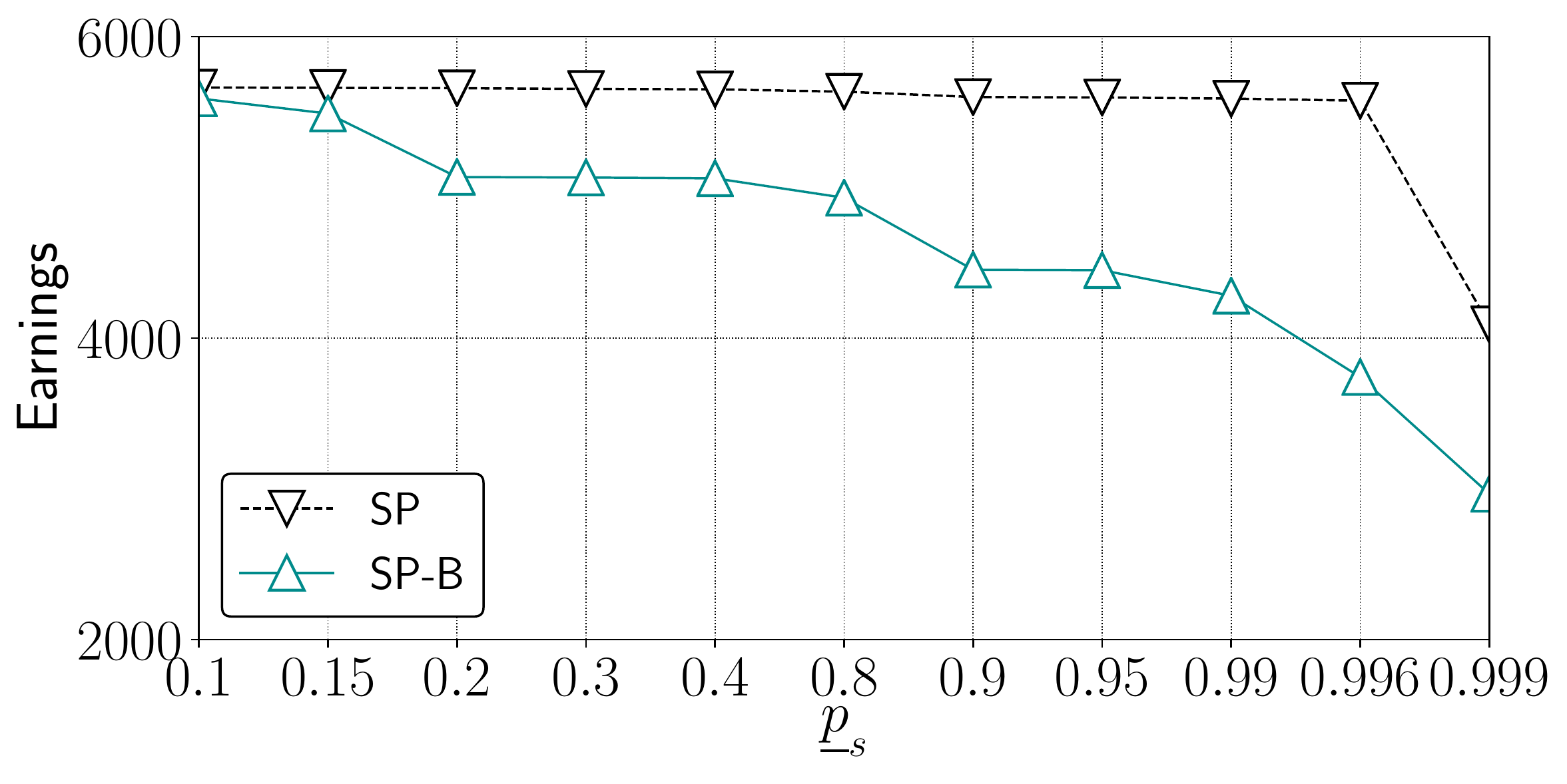}
\par\end{centering}
}
\par\end{centering}
\caption{Performance of the $\mathtt{SP\text{-}B}$ and $\mathtt{SP}$ approaches
on the provisioning of multiple slices of one type, with different
values of $\underline{p}_{s}$, in terms of (a) acceptance rate and
(b) total earnings.\label{fig:Eva_Multi1}}
\end{figure}

Now, considering $10$ slices of type~$1$ with the same $\underline{p}_{s}=0.99$
, the $\mathtt{SP\text{-}B}$ and $\mathtt{SP}$ variants are compared
in terms of acceptance rate, \textit{i.e}., percentage of slices that
have been successfully provisioned, given by $\sum_{s\in\mathcal{S}}\frac{x_{s}}{\left|\mathcal{S}\right|}$,
for different value of $\underline{p}_{s}$, see Figure~\ref{fig:Eva_Multi1_Accept_Rate}.
The tolerated impact probability $\overline{p}^{\text{im}}$ is set
to $0.1$. As expected, when $\underline{p}_{s}$ increases, the acceptance
rate decreases. Moreover, the $\mathtt{SP}$ approach, which does
not account for background services, has always a higher acceptance
rate compared to the $\mathtt{SP\text{-}B}$ approach.

\subsubsection{Provisioning of Several Slices of Different Types\label{subsec:Eva_Multi_3Types}}

The performance of the four variants is illustrated in this section,
when resources of $2$ to $8$ slices of three different types have
to be provisioned. The number of slices of each type and their associated
$\underline{p}_{s}$ are detailed in Table~\ref{tab:Scenarios}.
The impact probability threshold $\overline{p}^{\text{im}}$ is set
to $0.1$ in all scenarios.

\begin{table}[tbh]
\caption{Number of slices of each type as a function of $|\mathcal{S}|$\label{tab:Scenarios}}

\centering  
\begin{tabular}{    
p{0.1\columnwidth} 
p{0.1\columnwidth}    
p{0.1\columnwidth}         
p{0.1\columnwidth}
p{0.08\columnwidth}  
}  
\toprule  

\multicolumn{1}{c}{Case}& \multicolumn{1}{c}{\#Type 1}& \multicolumn{1}{c}{\#Type
2}& \multicolumn{1}{c}{ \#Type 3 }\\

\cmidrule[0.4pt](lr{0.12em}){1-1}%
\cmidrule[0.4pt](lr{0.12em}){2-2}%
\cmidrule[0.4pt](lr{0.12em}){3-3}%
\cmidrule[0.4pt](lr{0.12em}){4-4}%

\multicolumn{1}{c}{$|\mathcal{S}|=2$}&\multicolumn{1}{c}{ $1$
}& \multicolumn{1}{c}{ $1$ }& \multicolumn{1}{c}{ $0$ }\\

\multicolumn{1}{c}{$|\mathcal{S}|=4$}&\multicolumn{1}{c}{ $2$
}& \multicolumn{1}{c}{ $1$ }& \multicolumn{1}{c}{ $1$ }\\

\multicolumn{1}{c}{$|\mathcal{S}|=6$}&\multicolumn{1}{c}{ $2$
}& \multicolumn{1}{c}{ $2$ }& \multicolumn{1}{c}{ $2$ }\\

\multicolumn{1}{c}{$|\mathcal{S}|=8$}&\multicolumn{1}{c}{ $3$
}& \multicolumn{1}{c}{ $2$ }& \multicolumn{1}{c}{ $3$ }\\

\bottomrule
\end{tabular} 
\end{table}

\begin{figure}[H]
\begin{centering}
\subfloat[\label{fig:Eva_Multi2_PerNode}]{\begin{centering}
\includegraphics[width=0.23\textwidth]{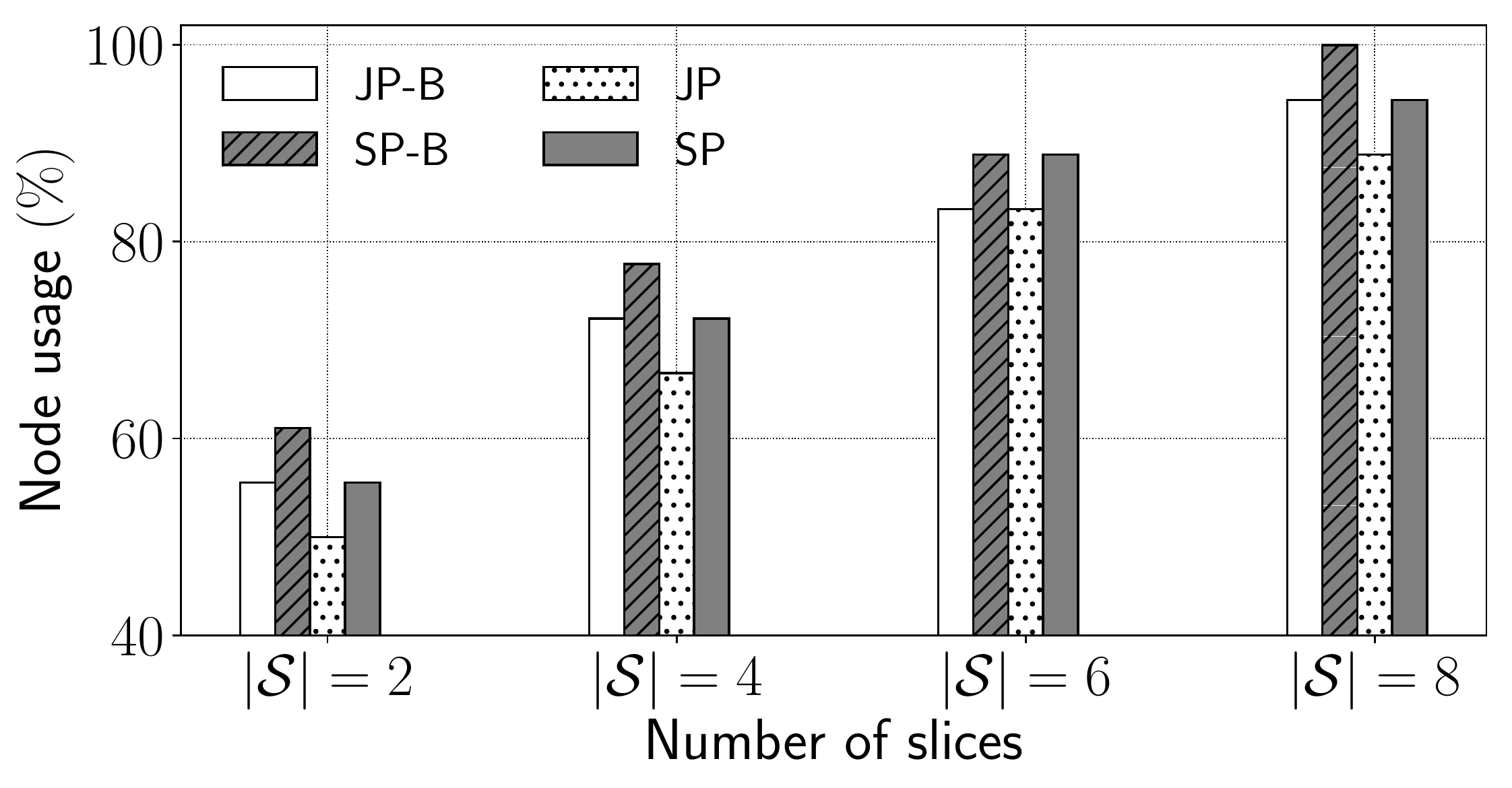}
\par\end{centering}
}\subfloat[\label{fig:Eva_Multi2_PerLink}]{\begin{centering}
\includegraphics[width=0.23\textwidth]{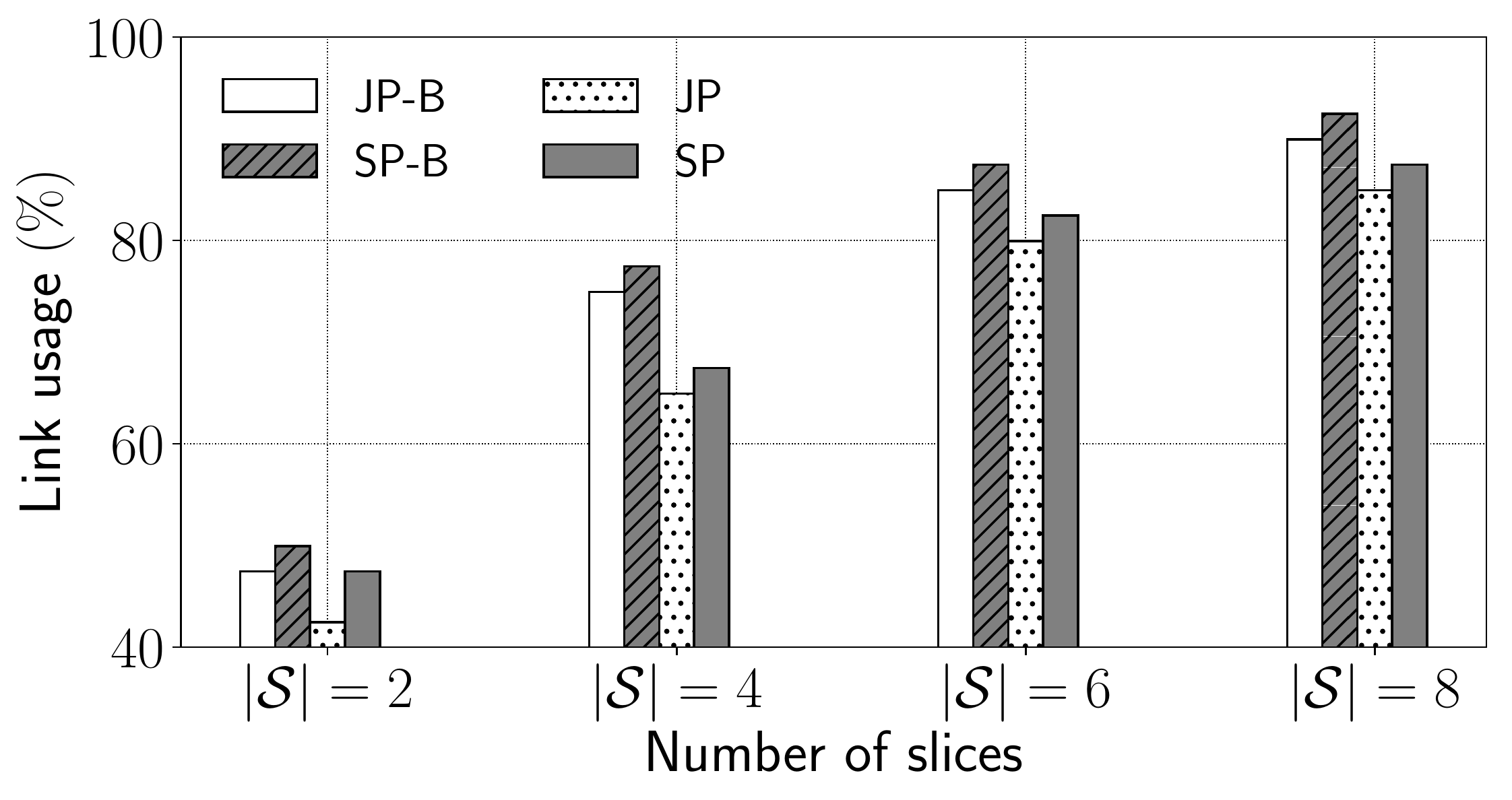}
\par\end{centering}
}
\par\end{centering}
\begin{centering}
\subfloat[\label{fig:Eva_Multi2_Cost}]{\begin{centering}
\includegraphics[width=0.23\textwidth]{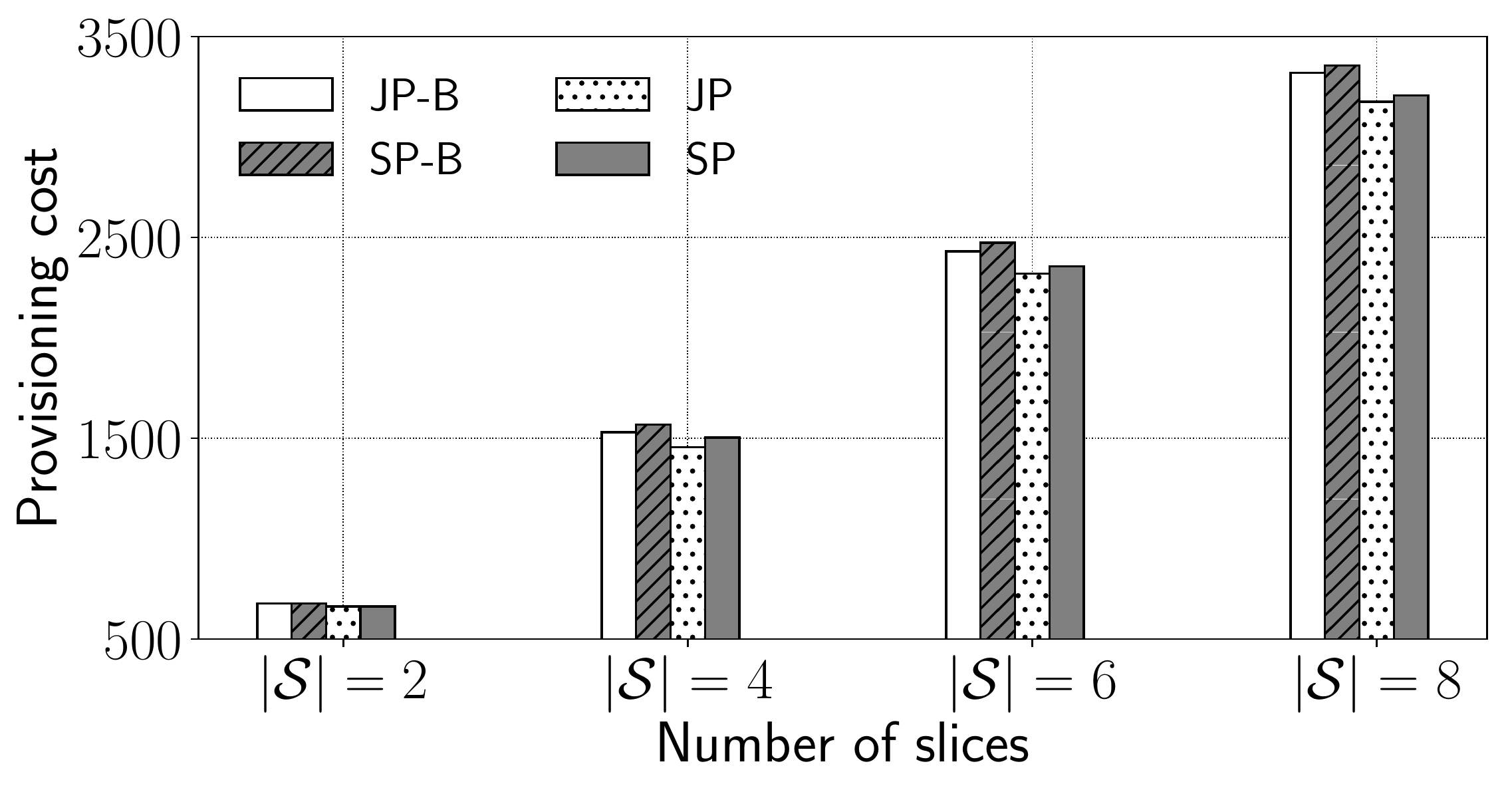}
\par\end{centering}
}\subfloat[\label{fig:Eva_Multi2_Earn}]{\begin{centering}
\includegraphics[width=0.23\textwidth]{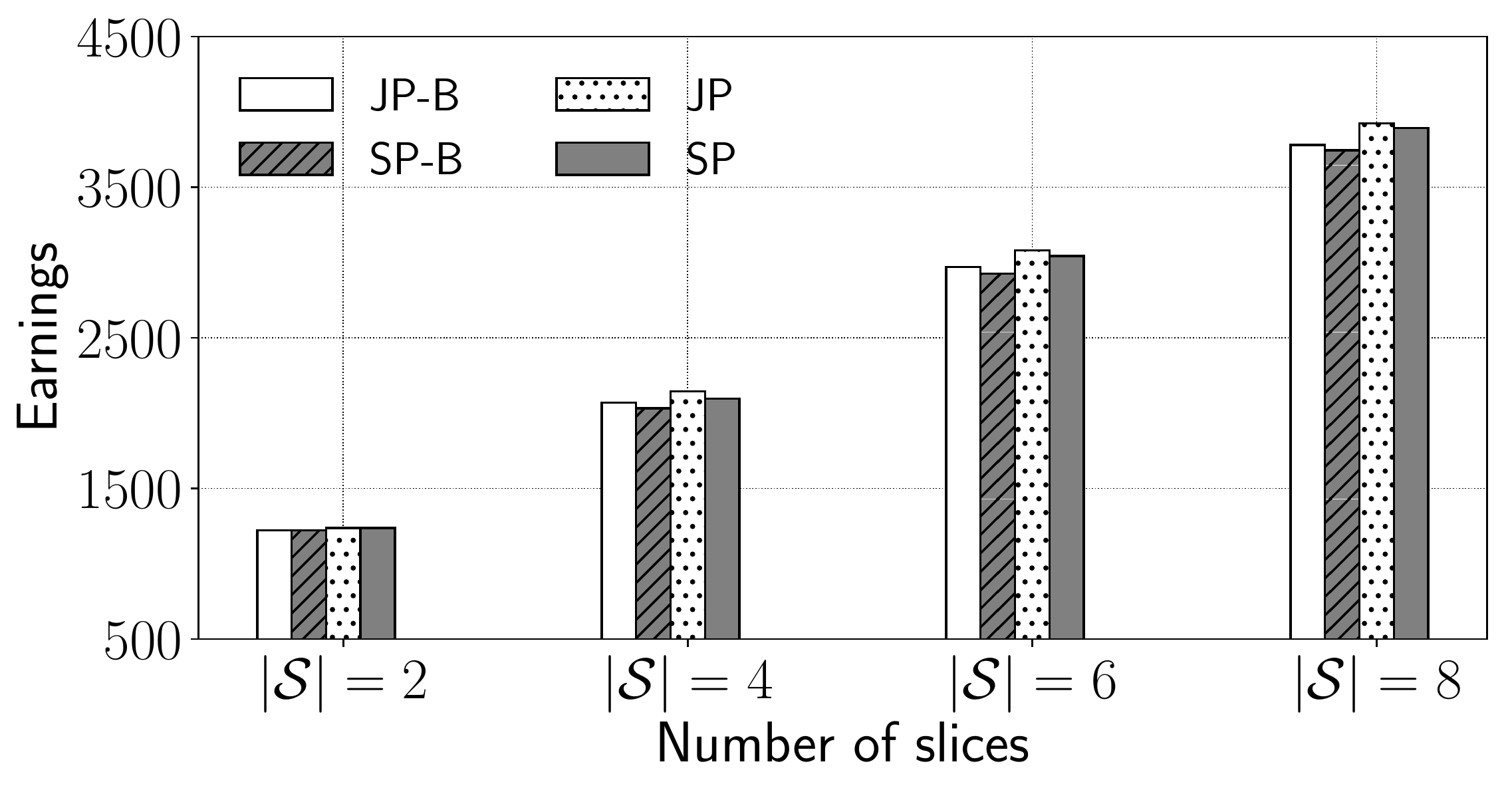}
\par\end{centering}
}
\par\end{centering}
\begin{centering}
\subfloat[\label{fig:Eva_Multi2_Nimpact}]{\begin{centering}
\includegraphics[width=0.228\textwidth]{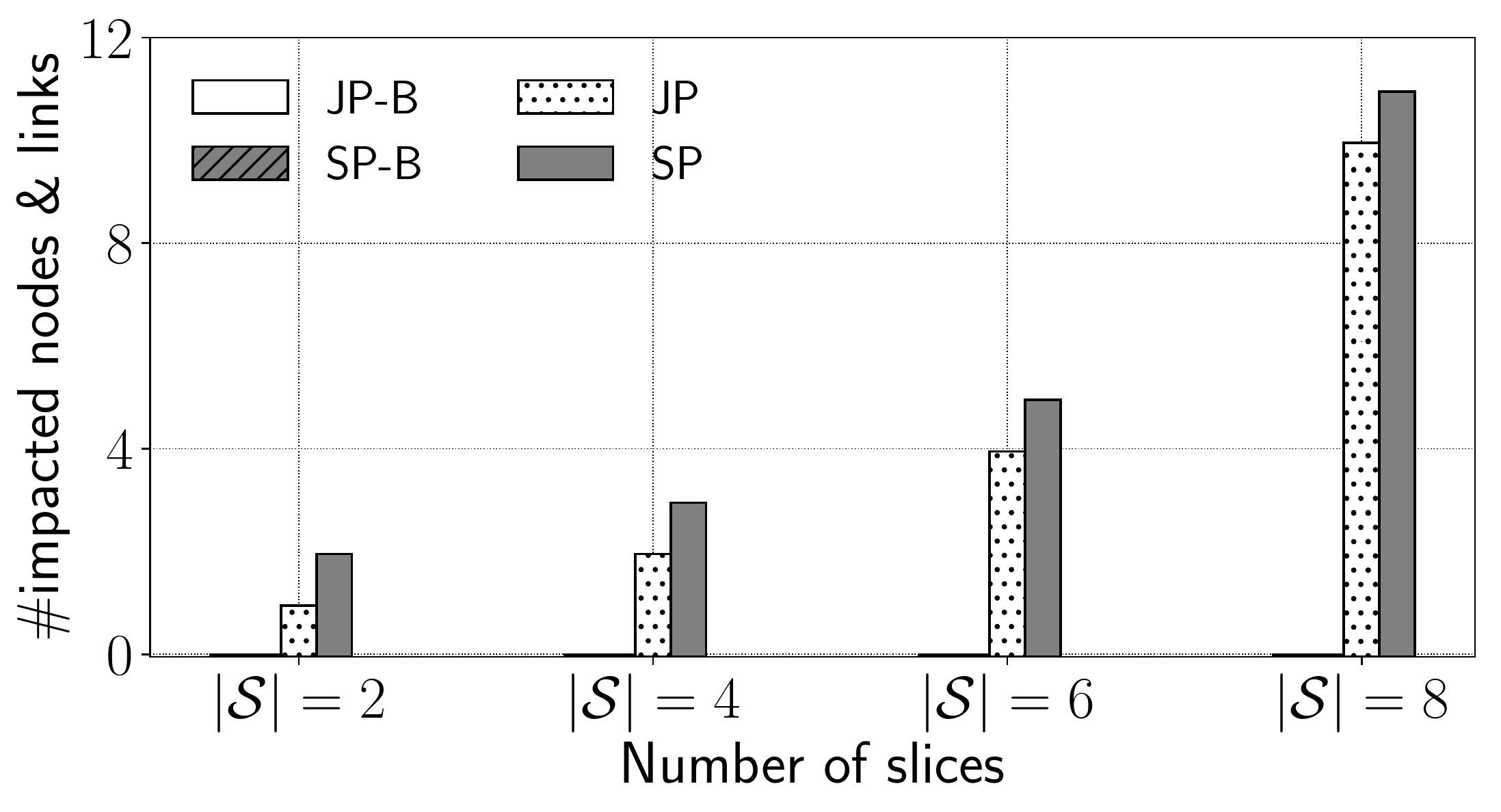}
\par\end{centering}
}\subfloat[\label{fig:Eva_Multi2_Time}]{\begin{centering}
\includegraphics[width=0.23\textwidth]{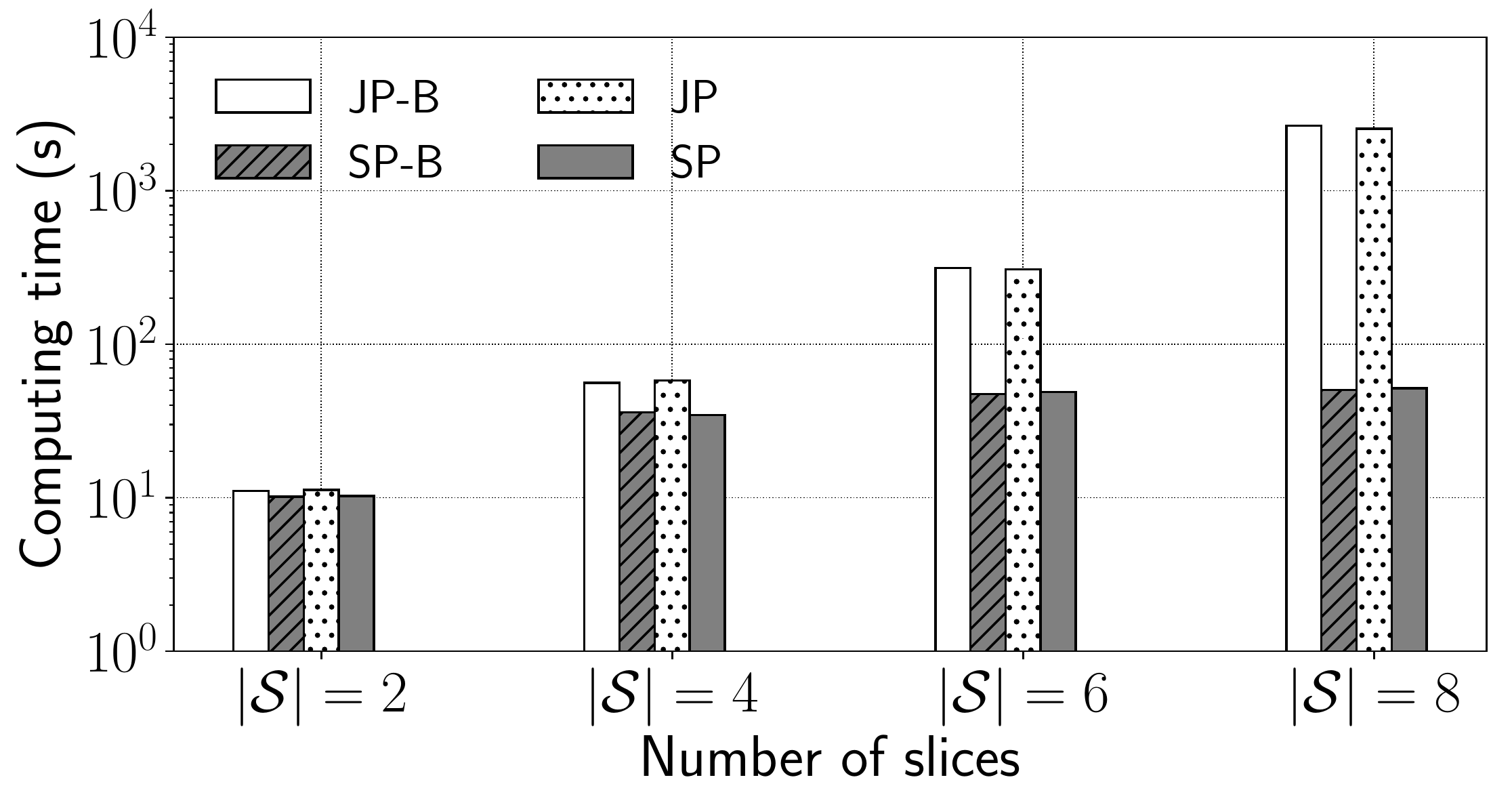}
\par\end{centering}
}
\par\end{centering}
\caption{Performance comparison of $4$ variants in terms of utilization of
infrastructure nodes (a), infrastructure links (b), provisioning costs
(c), total earnings (d), number of impacted nodes and links (e), and
computing time (f). \label{fig:Eva-Multi-Slices-Multi-Types}}
\end{figure}

The use of infrastructure nodes and links is shown in Figures~\ref{fig:Eva_Multi2_PerNode}
and \ref{fig:Eva_Multi2_PerLink}. The joint provisioning approaches
($\mathtt{JP}$ and $\mathtt{JP\text{-}B}$) require a reduced amount
of nodes and links compared to the sequential schemes ($\mathtt{SP}$
and $\mathtt{SP\text{-}B}$). Moreover, considering the impact on
background services requires, again, provisioning resources on more
nodes and links.

Figure~\ref{fig:Eva_Multi2_Cost} shows the provisioning costs obtained
with the various approaches. One observes that the $\mathtt{JP}$
variant yields the smallest cost among all variants, as it aims at
finding an optimal solution for all slices, without considering the
impact probability, contrary to the $\mathtt{JP\text{-}B}$ variant.
This leads to the highest earnings for the InP, as shown in Figure~\ref{fig:Eva_Multi2_Earn}.

The total number of impacted nodes and links is shown in Figure~\ref{fig:Eva_Multi2_Nimpact}.
The $\mathtt{JP\text{-}B}$ and $\mathtt{SP\text{-}B}$ variant have
no impacted nodes or links, whereas the provisioning performed by
the $\mathtt{JP}$ and $\mathtt{SP}$ approaches significantly impact
the background services. The $\mathtt{SP}$ variant has a higher impact
on the background services, due to the higher utilization of infrastructure
nodes and links, as shown in Figures~\ref{fig:Eva_Multi2_PerNode}
and \ref{fig:Eva_Multi2_PerLink}.

From the InP perspective, the use of impact-unaware variants ($\mathtt{JP}$
and $\mathtt{SP}$) maximizes its earning but violates background
services at a significant number of infrastructure nodes and links.
This may necessitate to reconfigure those background services. On
contrary, by using the impact-aware variants ($\mathtt{JP\text{-}B}$
and $\mathtt{SP\text{-}B}$), the InP can provision slices and preserve
a tolerable impact on the background services. The price to be paid
is somewhat degraded node and link utilization efficiency and a higher
provisioning cost compare to the impact-aware variants, leading to
a lower earnings for the InP. For instance, when provisioning for
$4$ slices, the $\mathtt{JP\text{-}B}$ variant uses around $72\%$
of the total infrastructure nodes to aggregate resources needed to
support the slices, while only $66.7\%$ of the nodes are employed
by the $\mathtt{JP}$ method, leading to a reduction of $3.5\%$ of
total earnings, as depicted in Figures~\ref{fig:Eva_Multi2_PerNode}
and \ref{fig:Eva_Multi2_Earn}.

As expected, the sequential provisioning methods ($\mathtt{SP\text{-}B}$
and $\mathtt{SP}$) perform better in terms of computing time than
the joint approaches ($\mathtt{JP\text{-}B}$ and $\mathtt{JP}$).
Increasing the number of slices leads to an increase of the cardinality
of the sets of variables $\boldsymbol{d}$ and $\boldsymbol{\kappa}$,
and therefore increases the computing time. In sequential provisioning,
slices are considered successively. There is only a very small difference
(usually less than $5\%$) in computing time between the $\mathtt{SP\text{-}B}$
and $\mathtt{SP}$ approaches and between the $\mathtt{JP\text{-}B}$
and $\mathtt{JP}$ approaches.

\begin{table*}[tbh]
\caption{Parameters of U-RD, SFC-RD, and S-RD graphs\label{tab:SRD}}

\centering
\scriptsize
\begin{tabular}{
p{0.05\columnwidth}
p{0.17\columnwidth}
p{0.17\columnwidth}
p{0.17\columnwidth}
p{0.19\columnwidth}
p{0.001\columnwidth} 
p{0.15\columnwidth}
p{0.14\columnwidth} 
p{0.05\columnwidth} 
}
\toprule

\multicolumn{8}{l}{\textbf{Type }1: HD video streaming at $4$~Mbps.
$N_{s}\sim\mathcal{B}\left(300,0.9\right)$, $I_{s}=900$, $\underline{p}_{s}=0.99$
}\\[5pt]

\multicolumn{1}{l}{ \textit{Node} }

& \multicolumn{1}{c}{ $\left(\mu_{s,\text{c}},\sigma_{s,\text{c}}\right)$
}

& \multicolumn{1}{c}{ $\left(\mu_{s,\text{m}},\sigma_{s,\text{m}}\right)$
}

& \multicolumn{1}{c}{ $\left(\mu_{s,\text{w}},\sigma_{s,\text{w}}\right)$
}

& \multicolumn{1}{c}{ $\left(r_{\text{c}},r_{\text{m}},r_{\text{w}}\right)$
}

&& \multicolumn{1}{c}{ \textit{Link} }

& \multicolumn{1}{c}{ $\left(\mu_{\text{b}},\sigma_{\text{b}}\right)$
}

& \multicolumn{1}{c}{ $r_{s,\text{b}}$ }\\

\cmidrule[0.4pt](lr{0.12em}){1-1}%
\cmidrule[0.4pt](lr{0.12em}){2-2}%
\cmidrule[0.4pt](lr{0.12em}){3-3}%
\cmidrule[0.4pt](lr{0.12em}){4-4}%
\cmidrule[0.4pt](lr{0.12em}){5-5}%
\cmidrule[0.4pt](lr{0.12em}){7-7}%
\cmidrule[0.4pt](lr{0.12em}){8-8}%
\cmidrule[0.4pt](lr{0.12em}){9-9}%

vVOC& $\left(5.4,0.54\right)\textrm{e-}3$ &$\left(1.5,0.15\right)\textrm{e-}2$
&\multicolumn{1}{c}{\textemdash }&$\left(0.29,0.81,0\right)$&&
vVOC$\rightarrow$vGW & $\left(4,0.4\right)\textrm{e-}3$ & $0.22$\\

vGW&$\left(9.0,0.90\right)\textrm{e-}4$&$\left(5.0,0.50\right)\textrm{e-}4$&\multicolumn{1}{c}{\textemdash }&$\left(0.05,0.03,0\right)$&&vGW$\rightarrow$vBBU&$\left(4,0.4\right)\textrm{e-}3$&$0.22$\\

vBBU&$\left(8.0,0.80\right)\textrm{e-}4$&$\left(5.0,0.50\right)\textrm{e-}4$&$\left(4,0.4\right)\textrm{e-}3$
&$\left(0.04,0.03,0.2\right)$ && &&\\[10pt]

\multicolumn{8}{l}{\textbf{Type 2}: SD video streaming at $2$~Mbps.
$N_{s}\sim\mathcal{B}\left(1000,0.8\right)$, $I_{s}=1000$, $\underline{p}_{s}=0.95$
}\\[5pt]

\multicolumn{1}{l}{ \textit{Node} }

& \multicolumn{1}{c}{ $\left(\mu_{s,\text{c}},\sigma_{s,\text{c}}\right)$
}

& \multicolumn{1}{c}{ $\left(\mu_{s,\text{m}},\sigma_{s,\text{m}}\right)$
}

& \multicolumn{1}{c}{ $\left(\mu_{s,\text{w}},\sigma_{s,\text{w}}\right)$
}

& \multicolumn{1}{c}{ $\left(r_{\text{c}},r_{\text{m}},r_{\text{w}}\right)$
}

&& \multicolumn{1}{c}{ \textit{Link} }

& \multicolumn{1}{c}{ $\left(\mu_{\text{b}},\sigma_{\text{b}}\right)$
}

& \multicolumn{1}{c}{ $r_{s,\text{b}}$ }\\

\cmidrule[0.4pt](lr{0.12em}){1-1}%
\cmidrule[0.4pt](lr{0.12em}){2-2}%
\cmidrule[0.4pt](lr{0.12em}){3-3}%
\cmidrule[0.4pt](lr{0.12em}){4-4}%
\cmidrule[0.4pt](lr{0.12em}){5-5}%
\cmidrule[0.4pt](lr{0.12em}){7-7}%
\cmidrule[0.4pt](lr{0.12em}){8-8}%
\cmidrule[0.4pt](lr{0.12em}){9-9}%

vVOC &$\left(1.1,0.11\right)\textrm{e-}3$&$\left(7.5,0.75\right)\textrm{e-}3$&\multicolumn{1}{c}{\textemdash }&$\left(0.17,1.20,0\right)$&&
vVOC$\rightarrow$vGW &$\left(2,0.2\right)\textrm{e-}3$ &$0.32$\\

vGW &$\left(1.8,0.18\right)\textrm{e-}4$&$\left(2.5,0.25\right)\textrm{e-}4$&\multicolumn{1}{c}{\textemdash }&$\left(0.03,0.04,0\right)$&&
vGW$\rightarrow$vBBU &$\left(2,0.2\right)\textrm{e-}3$&$0.32$\\

vBBU &$\left(0.8,0.08\right)\textrm{e-}4$&$\left(2.5,0.25\right)\textrm{e-}4$&$\left(2,0.2\right)\textrm{e-}3$
&$\left(0.01,0.04,0.3\right)$ &&&&\\[10pt]

\multicolumn{7}{l}{\textbf{Type 3}: Video surveillance and traffic
monitoring at $1$~Mbps. $N_{s}=50$, $I_{s}=800$, $\underline{p}_{s}=0.9$
}\\[5pt]

\multicolumn{1}{l}{ \textit{Node} }

& \multicolumn{1}{c}{ $\left(\mu_{s,\text{c}},\sigma_{s,\text{c}}\right)$
}

& \multicolumn{1}{c}{ $\left(\mu_{s,\text{m}},\sigma_{s,\text{m}}\right)$
}

& \multicolumn{1}{c}{ $\left(\mu_{s,\text{w}},\sigma_{s,\text{w}}\right)$
}

& \multicolumn{1}{c}{ $\left(r_{\text{c}},r_{\text{m}},r_{\text{w}}\right)$
}

&& \multicolumn{1}{c}{ \textit{Link} }

& \multicolumn{1}{c}{ $\left(\mu_{\text{b}},\sigma_{\text{b}}\right)$
}

& \multicolumn{1}{c}{ $r_{s,\text{b}}$ }\\

\cmidrule[0.4pt](lr{0.12em}){1-1}%
\cmidrule[0.4pt](lr{0.12em}){2-2}%
\cmidrule[0.4pt](lr{0.12em}){3-3}%
\cmidrule[0.4pt](lr{0.12em}){4-4}%
\cmidrule[0.4pt](lr{0.12em}){5-5}%
\cmidrule[0.4pt](lr{0.12em}){7-7}%
\cmidrule[0.4pt](lr{0.12em}){8-8}%
\cmidrule[0.4pt](lr{0.12em}){9-9}%

vBBU&$\left(2.0,0.20\right)\textrm{e-}4$&$\left(1.3,0.13\right)\textrm{e-}4$&$\left(1,0.1\right)\textrm{e-}3$&$\left(0.4,0.25,2\right)\textrm{e-}2$&&
vBBU$\rightarrow$vGW&$\left(1,0.1\right)\textrm{e-}3$&$0.02$\\

vGW&$\left(9.0,0.90\right)\textrm{e-}4$&$\left(1.3,0.13\right)\textrm{e-}4$&\multicolumn{1}{c}{\textemdash }&$\left(0.018,0.003,0\right)$
&&vGW$\rightarrow$vTM&$\left(1,0.1\right)\textrm{e-}3$&$0.02$\\

vTM&$\left(1.1,0.11\right)\textrm{e-}3$&$\left(1.3,0.13\right)\textrm{e-}4$&\multicolumn{1}{c}{\textemdash }&$\left(0.266,0.003,0\right)$
&&vTM$\rightarrow$vVOC&$\left(1,0.1\right)\textrm{e-}3$&$0.02$\\

vVOC&$\left(5.4,0.54\right)\textrm{e-}3$&$\left(3.8,0.38\right)\textrm{e-}3$&\multicolumn{1}{c}{\textemdash }&$\left(0.108,0.080,0\right)$&&vVOC$\rightarrow$vIDPS&$\left(1,0.1\right)\textrm{e-}3$&$0.02$\\

vIDPS&$\left(1.1,0.11\right)\textrm{e-}2$&$\left(1.3,0.13\right)\textrm{e-}4$&\multicolumn{1}{c}{\textemdash }&$\left(0.214,0.003,0\right)$&&&&\\

\bottomrule  
\end{tabular} 
\end{table*}

\section{Conclusions\label{sec:Conclusions}}

This paper investigates a resource provisioning method for network
slicing robust to a partly unknown number of users whose resource
demands are uncertain. Adopting the point of view of the InP, one
tries to maximize its earnings, while providing a probabilistic guarantee
that the slice resource demands are fulfilled. In addition to that,
the proposed resource provisioning method is performed to keep the
impact on the background services under a threshold imposed by the
InP.

The uncertainty-aware slice resource provisioning is formulated as
a nonlinear constrained optimization problem. A parameterized MILP
formulation is then proposed. With the MILP formulation, four variants
($\mathtt{JP}$, $\mathtt{SP}$, $\mathtt{JP\text{-}B}$, and $\mathtt{SP\text{-}B}$)
are introduced, for the solution of the provisioning problem for multiple
slices jointly or sequentially, without or with consideration of the
impact on background services.

The \textit{\emph{impact-limiting}} variants ($\mathtt{JP\text{-}B}$,
and $\mathtt{SP\text{-}B}$) have a controlled impact on the background
services, whereas the $\mathtt{JP}$ and $\mathtt{SP}$ variants,
which do not care of the impact on background services, consume all
resources of several infrastructure nodes and links, which may impose
a reconfiguration of background services. The price to be paid for
the InP with impact-limiting variants are lower earnings.

Moreover, due to the exponential worst-case complexity in the number
of variables of the MILP formulation, as expected, sequential approaches
are shown to better scale to a larger number of slices. The price
to be paid by the sequential approaches is a somewhat degraded node
and link utilization, a higher provisioning cost, and lower earnings,
compared to the joint approaches. 

In this paper, uncertainties related to the fluctuation of user demands
and the background services have been taken into account for the slice
resource provisioning. A prospective extension to this work is to
let the InP, if necessary, update the already provisioned resources
for some slices during their lifetime. This allows one to have a more
realistic adaptive SLAs and dynamic provisioning techniques for network
slicing.

\vskip -2\baselineskip plus -1fil
\bibliographystyle{IEEEtran} 
\bibliography{ref_uncertain}

\end{document}